\documentclass[useAMS,usenatbib]{mn2e}
\usepackage{deluxetable}
\usepackage{epsf}
\usepackage{amsmath}
\usepackage{subfigure}
\usepackage{url}
\usepackage{color}

\newcommand{\lya}{Ly$\alpha$\ }
\newcommand{\hkpc}{h^{-1}{\rm kpc}}
\newcommand{\hmpc}{h^{-1}{\rm Mpc}}

\newcommand{\msun}{M_{\odot}}

\newcommand{\dtr}{\Delta_{\rm{thr}}}
\newcommand{\dlls}{\Delta_{\rm{lls}}}
\newcommand{\Ghi}{\Gamma_{\rm{HI}}}

\newcommand{\aap}{A\&A}
\newcommand{\apjs}{ApJS}
\newcommand{\apj}{ApJ}
\newcommand{\apjl}{ApJL}
\newcommand{\aj}{AJ}

\newcommand{\mnras}{MNRAS}

\newcommand{\cnenp}{C_{{\rm H\,II}}}
\newcommand{\cnenptb}{C_{{\rm H\,II,T_b}}}
\newcommand{\cnenpobs}{C_{{\rm H\,II,}10^4{\rm K}}}
\newcommand{\cb}{C_{\mathrm{b}}}
\newcommand{\fesc}{f_{\mathrm{esc}}}
\newcommand{\fescfive}{f_{\mathrm{esc,5}}}

\newcommand{\nhi}{n_{\mathrm{H\,I}}}
\newcommand{\nh}{n_{\mathrm{H}}}
\newcommand{\nem}{n_{\mathrm{e}}}

\newcommand{\nhii}{n_{\mathrm{H\,II}}}

\newcommand{\fJ}{f_{\mathcal{J}}}

\newcommand{\tes}{\tau_{\mathrm{es}}}
\newcommand{\ta}{\tau_\alpha}

\newcommand{\xhim}{x_{\mathrm{H\,I,M}}}
\newcommand{\xhi}{x_{\mathrm{H\,I}}}
\newcommand{\xhii}{x_{\mathrm{H\,II}}}
\newcommand{\xhiv}{x_{\mathrm{H\,I,V}}}
\newcommand{\xhiiv}{x_{\mathrm{H\,II,V}}}

\newcommand{\dxrt}{\Delta x_{\mathrm{RT}}}

\begin{document}

\title[Gas Clumping at $z\geq5$]{Gas Clumping in Self-Consistent Reionisation Models}

\author[K.\ Finlator et al.]{
\parbox[t]{\textwidth}{\vspace{-1cm}
Kristian Finlator$^{1\dagger}$, S.\ Peng Oh$^1$, Feryal \"{O}zel$^2$, Romeel Dav\'e$^2$}
\\\\$^1$Department of Physics, University of California, Santa Barbara, CA
93106, USA
\\$^\dagger$Hubble Fellow
\\$^2$Astronomy Department, University of Arizona, Tucson, AZ 85721, USA
\author[Gas Clumping at $z\geq5$]{K.\ Finlator, S.\ Peng Oh, F. \"{O}zel, \& Romeel Dav\'e}
}

\maketitle

\begin{abstract}
We use a suite of cosmological hydrodynamic simulations including a self-consistent
treatment for inhomogeneous reionisation to study the impact of galactic outflows
and photoionisation heating on the volume-averaged recombination rate of the 
intergalactic medium (IGM).  By
incorporating an evolving ionising escape fraction and a treatment for 
self-shielding within Lyman limit systems, we have run the first simulations of 
``photon-starved" reionisation scenarios
that simultaneously reproduce observations of the abundance of galaxies, the
optical depth to electron scattering of cosmic microwave background photons 
$\tes$, and the effective optical depth to \lya absorption at $z=5$.  We 
confirm that an ionising background reduces the clumping factor $C$ by more than 50\% by 
smoothing moderately-overdense ($\Delta=$1--100) regions.  Meanwhile, outflows 
increase clumping only modestly.  The clumping factor of ionised gas is much 
lower than the overall baryonic clumping factor because the most overdense gas is 
self-shielded.  Photoionisation heating further suppresses recombinations if 
reionisation heats gas above the canonical 10,000 K.  Accounting for both 
effects within our most realistic simulation, $C$ rises from $<1$ at $z>10$ 
to 3.3 at $z=6$.  We show that incorporating temperature- and 
ionisation-corrected clumping factors into an analytical reionisation model 
reproduces the numerical simulation's $\tes$ to within 10\%.  Finally, we 
explore how many ionising photons are absorbed during the process of 
heating filaments by considering the overall 
photon cost of reionisation in analytical models that assume that the IGM is 
heated at different redshifts.  For reionisation redshifts of 9--10, 
cold filaments boost the reionisation photon budget by $\sim1$ photon 
per hydrogen atom.
\end{abstract}

\begin{keywords}
cosmology: theory --- radiative transfer --- hydrodynamics --- methods: numerical
\end{keywords}

\section{Introduction} \label{sec:intro}
Much of the current interest in cosmological hydrogen 
reionisation may be distilled to two related questions: (1) When did it 
occur?; and (2) What sources dominated the ionising photon budget?
Theoretical efforts to address the first question have traditionally 
considered two observational constraints: the optical depth to Thomson 
scattering of cosmic microwave background (CMB) photons, $\tes$, and the 
opacity of the Lyman-$\alpha$ forest~\citep[][and references therein]{tra09}.
While these works have shed much light on the nature of inhomogeneous
reionisation, their predictions have until recently been uncertain owing
to the unknown relation between ionising luminosity and halo 
mass~\citep{mcq07}.  With the recent 
installation of the Wide-Field Camera 3 on board the Hubble Space 
Telescope, a new generation of observations is now constraining the 
abundance and colors of galaxies back to $z=8$ and 
beyond~\citep{fin10,bou11,dun12,gon11,gra11,mcl11,oes12}.  These 
observations may be used to constrain the relationship between star 
formation rate and halo mass via abundance-matching 
studies~\citep{tre10,mun11}.  Furthermore, the relationship between ionising 
luminosity and star formation rate is now constrained by the direct 
detection of escaping Lyman continuum flux from high-redshift 
galaxies~\citep{sha06,sia10,nes11}.  Hence there are now tenuous observational 
links connecting halo mass with ionising luminosity, a major step toward
understanding the relationship between high-redshift galaxies and their 
environment.  

By combining observations of galaxies, the CMB, and the Lyman-$\alpha$ 
forest, it is possible to ask whether the observed ionising 
sources and reionisation history are 
compatible~\citep[for example,][]{haa12,kuh12}.  
Two uncertainties hinder progress toward answering this question.  
The first remains the fraction of ionising photons $\fesc$ that escape 
into the IGM: direct constraints are only available at lower 
redshifts and higher luminosities than are relevant to reionisation.  
Furthermore, $\fesc$ is sensitive to parsec-scale physical processes 
that occur deep within galaxies' interstellar media and is difficult 
to model numerically~\citep[][and references therein]{fer11}.  The second 
uncertainty involves the recombination rate of the reionisation-epoch 
IGM.  The unknown ratio of the IGM's true recombination rate to its hypothetical 
rate under the assumption of uniform density and temperature is often referred 
to as the IGM clumping factor $C$.  We will define $C$ explicitly in
Section~\ref{sec:defs}; for the present the important point is that it is
proportional to the volume-averaged hydrogen recombination rate in the IGM.
High values of $C$ require abundant faint galaxies and/or a large $\fesc$ 
while low values of $C$ are easier to reconcile with existing observations.  
$C$ is sensitive to the IGM's density, ionisation, and temperature fields, 
hence an accurate estimate requires three-dimensional numerical simulations.  

\begin{figure}
\centerline{
\setlength{\epsfxsize}{0.5\textwidth}
\centerline{\epsfbox{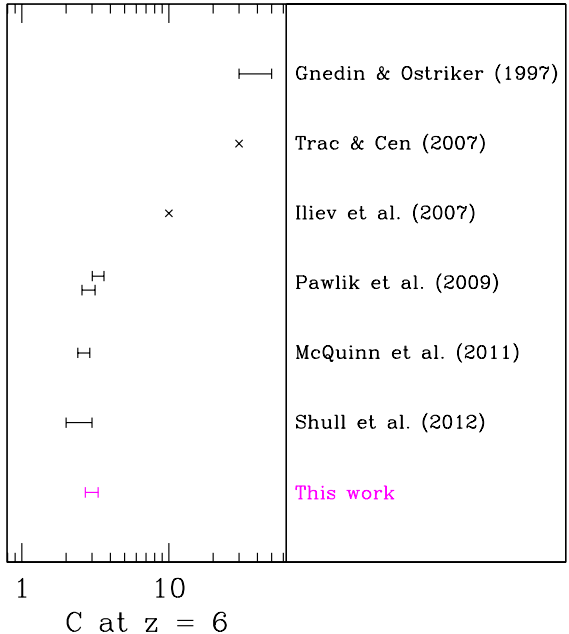}}
}
\caption{Our ionisation- and temperature-corrected IGM clumping 
factor at $z=6$ in comparison to previous calculations.  Our work 
indicates that the clumping factor lies within the range 2.7--3.3, 
which is consistent with the low values that have been found
in recent works.
}
\label{fig:clumplit}
\end{figure}

To this end, $C$ has been studied numerous times using numerical 
simulations.  We compile predictions of the clumping factor at $z=6$ 
from a variety of works in Figure~\ref{fig:clumplit}.
The early work of~\citet{gne97} yielded a clumping factor
of 30--50 (their Figure 2) depending on the choice of definition.  However,
this clumping factor included the recombinations that occur in gas whose 
ionisation state is dominated by the local field rather than by the global 
extragalactic ultraviolet ionising background (EUVB), hence it is an 
overestimate of the recombination rate within the gas 
that is conventionally associated with the IGM (see 
also~\citealt{gne00} and~\citealt{mcq11}).  For the same 
reason,~\citet{tra07} also inferred a clumping factor of roughly 30 
from simulations that subtended a significantly larger cosmological 
volume.

\citet{ili07} applied this insight to calculate an improved clumping 
factor using high-resolution N-body simulations whose spatial resolution 
was comparable to the Jeans length in the IGM.  By excluding matter within 
virialized halos, they derived a redshift-dependent fitting function that 
climbs to $\approx10$ at $z=6$, a value which was also found in the recent
N-body simulations of~\citet{rai11}.  This work confirmed that 
physically-motivated
density cuts reduce $C$.  However, as the 
authors noted, their calculation did not account for the fact that 
photoionisation heating (photoheating) tends to smooth the diffuse IGM.  
We will confirm 
in Figure~\ref{fig:pdf} that this leads to an overestimated clumping factor.

\citet{paw09a} corrected this problem by running cosmological hydrodynamic
simulations that model reionisation using a spatially-homogeneous EUVB.
They used several plausible density cuts to calculate the IGM clumping 
factor as a function of redshift.  We show their predictions assuming that
the IGM corresponds to overdensity cuts of 
$\Delta \equiv \rho/\langle \rho \rangle < 50$ (lower range) 
and $\Delta < 100$ (upper range).  This work generally confirmed the
impression that considering only low-density gas and accounting for
the tendency for the photoheating of diffuse regions 
both reduce $C$.  However, it suffered
from two drawbacks.  First, their use of a spatially-homogeneous EUVB
pressurized regions that should in reality be self-shielded, leading 
to uncertainty in the gas density distribution from which the 
predicted clumping factors were derived.  Second, their analysis
did not account explicitly for the IGM temperature, which could 
suppress the recombination rate further. 

In a qualitatively similar work,~\citet{shu12} ran a cosmological
hydrodynamic simulation with a spatially-homogeneous EUVB and 
computed the temperature-corrected clumping factor within gas 
that satisfied cuts on baryon overdensity, temperature, metallicity, 
and ionisation state.  Although these extra cuts render direct 
comparisons with~\citet{paw09a} uncertain, the inferred range is 
consistent.  Both works are also in reasonable agreement 
with~\citet{fau08}, who used the gas density distribution 
of~\citet{mir00} to find that $C=$3--4 if the clumping factor is
averaged only over gas with overdensity $\Delta$ less than 67.

The uncertainty in the derived clumping factor may be reduced by 
using radiation transport simulations to identify directly the gas 
that is self-shielded.  ~~\citet{mcq11} use this approach in 
post-processing to derive a clumping factor of 2.4--2.9.  Their
calculations do not account for the IGM temperature and,
like the works of~\citet{paw09a} and~\citet{shu12}, begin with
hydrodynamic simulations in which the moderately-overdense gas is 
incorrectly exposed to the EUVB (although they confirm directly 
that shielding gas that is dense enough to form stars has no
impact).  Nonetheless, their results are clearly consistent with
a relatively low clumping factor.

The goal of the present work is to combine all of these ideas into a 
single radiation hydrodynamic framework that accounts realistically 
for the time and spatial dependence of the IGM's density, ionisation, 
and temperature fields.  Our fiducial simulation reproduces the
observed UV luminosity function of galaxies at 
$z\geq6$~\citep{fin11} and is marginally consistent with the effective 
optical depth to absorption of \lya photons and $\tes$.  Hence, our 
estimated IGM recombination rate is constrained by a variety of observations.  
As Figure~\ref{fig:clumplit} indicates, we will confirm that effects 
that have been neglected in recent works based on hydrodynamic 
simulations such as incorrectly heating dense regions and ignoring 
the gas temperature are indeed small.  Of course, the 
purpose of clumping factors is to inform analytical models of 
reionisation~\citep[for example,][]{mad99}.
We will therefore verify that analytical reionisation calculations 
incorporating clumping factors do, in fact, reproduce the behavior 
of numerical simulations.  This check will assess directly the 
magnitude of the systematic uncertainties inherent in calculations 
that ignore higher-order effects such as shadowing, source 
clustering, and the light-travel time of ionising photons.

Having verified that our analytical model reproduces the behavior of
our numerical model reasonably well, we will use it to ask how
many photons are absorbed by filamentary overdensities in the 
diffuse IGM.  This will quantify the error introduced in models
that compute reionisation using IGM density distributions in 
which the gas is effectively preheated prior to the onset of 
reionisation.

The outline of our paper is as follows: We begin by introducing our
suite of simulations in Section~\ref{sec:sims}.  We compare
different definitions of the clumping factor in Section~\ref{sec:defs}
and explore how feedback effects modulate their values in 
Section~\ref{sec:feedback}.  In Section~\ref{sec:photCount}, we
incorporate our derived clumping factors into an analytical model
for reionisation and test how well analytical models of
reionisation reproduce the behavior of numerical calculations.
We apply this model to study how many photons are absorbed in
filaments in Section~\ref{sec:photCost}.  Finally, we discuss 
our results in Section~\ref{sec:disc} and summarize in 
Section~\ref{sec:sum}.  We introduce two improvements to our
treatment of the Eddington tensors in Appendix~\ref{app:fedd}
and quantify resolution limitations in Appendix~\ref{app:res}. 

\section{Simulations} \label{sec:sims}

\subsection{Star Formation and Outflows}
Our numerical methods, input physics, and cosmology are similar 
to~\citet{fin11} with the exception of our adopted treatments
for an evolving $\fesc$, subgrid self-shielding (see below),
and two numerical optimizations regarding the treatment of 
the Eddington tensor field (Appendix~\ref{app:fedd}). 
We model the growth of structure and the feedback processes 
that couple galaxies with the IGM using our custom version of the
cosmological galaxy formation code {\sc Gadget-2}~\citep{spr05}.
{\sc Gadget-2} implements a formulation of smoothed particle 
hydrodynamics (SPH) that simultaneously conserves entropy and 
energy and solves for the gravitational potential with a 
tree-particle-mesh algorithm.  Dense gas cools radiatively
using the primordial cooling processes in Table 1 of~\citet{kat96}
and the metal-line cooling tables of~\citet{sut93}, which assume
collisional ionisation equilibrium.  In contrast with~\citet{kat96},
however, we evolve the ionisation states of hydrogen and helium and 
the electron abundance simultaneously with the cooling equations 
using a nested subcycling approach whose timestep is limited by a 
chemical Courant condition; see~\citet{fin11} for details.  
We initialize the IGM temperature and neutral hydrogen fraction to 
the values appropriate for each simulation's initial redshift as 
computed by {\sc recfast}~\citep{won08}, and we assume that 
helium is initially completely neutral.  We generate the initial 
density field using an~\citet{eis99} power spectrum at redshifts 
of 249 and 319 for simulations subtending 6 and 3 $\hmpc$, 
respectively.  All simulations assume a cosmology in which 
$\Omega_M=0.28$, $\Omega_\Lambda=0.72$, $\Omega_b = 0.046$, 
$h=0.7$, $\sigma_8 = 0.82$, and the index of the primordial power 
spectrum $n=0.96$.

Our goal of modeling galaxies and reionisation simultaneously
requires a treatment for the ability of cool gas to form stars.
We adopt the subgrid two-phase interstellar medium treatment
of~\citet{spr03}, which can be tuned to reproduce the observed
relation between the surface densities of gas and star 
formation~\citep{ken98}.  The physical density threshold for star
formation is 0.13 cm$^{-3}$.  This value is motivated by observations
of a critical density for the onset of star formation and lies within
the range at which the thermo-gravitational instability is expected
to become active~\citep{sch04}.  Varying it has only a minor impact
on the predicted star formation rate density~\citep{sch10}.
We account for metal enrichment owing
to supernovae of Types II and Ia as well as asymptotic giant branch
stars; see~\citet{opp08} for details.

Allowing cold gas to form stars without any feedback invariably 
results in overproducing the observed reionisation-epoch star formation 
rate density~\citep{dav06,fin11}.  The accepted solution is to allow 
feedback from massive stars to expel star-forming gas from galaxies.
As the spatial resolution necessary to form such outflows 
self-consistently is beyond the reach of current cosmological 
simulations~\citep{mac99,hop12,pow11}, we impose star formation-driven 
outflows by stochastically applying kicks to star-forming gas 
particles.  The amount of gas kicked per unit stellar mass formed 
and the kick velocities are adjusted to reproduce the expected 
scalings from momentum-driven outflows~\citep{mur05}.  We temporarily 
disable hydrodynamic forces in outflowing gas in order to
mimic the way in which outflows escape through holes in
higher-resolution simulations~\citep{mac99}.  Hydrodynamic forces are 
restored once the gas has traveled for $1.95\times10^{4}$ km s$^{-1}/v$ 
million years (where $v$ is the kick velocity) or its density drops
below 10\% of the threshold density for star formation.  In practice, this
allows outflowing gas to reach galactocentric radii of 50--100 
physical kpc and then re-accrete onto the central galaxy on a timescale 
of $\sim1$ Gyr~\citep{opp08}.  The distance that outflows travel before 
re-accreting is large compared to the typical radius at which hydrodynamic
forces are enabled, hence the decoupling prescription has little impact
on outflows once they escape the ISM.  The long re-accretion timescale enables
expelled gas to modify the IGM recombination rate and opacity depending
on the density of outflowing gas; exploring these possibilities is one 
of the goals of the present work.

\subsection{Radiation Transport}\label{ssec:rt}
We evolve the EUVB on a fixed Cartesian grid using the 
moments of the cosmological radiative transfer equation.  We close
the moment hierarchy with a variable Eddington tensor field that is
updated periodically via a time-independent ray-casting calculation.
This approach accounts accurately for the inhomogeneous opacity 
field as long as the updates occur frequently 
enough~\citep{fin09a}.  Each cell's ionising luminosity is given by
a sum over the luminosities of its star-forming gas particles. 
Their luminosities, in turn, depend on their star formation rates
and metallicities, where the metallicity dependence follows an 
analytical fit to Table 4 of~\citet{sch03}.  

The fraction of 
ionising photons that escapes into the IGM in order to participate in
reionisation $\fesc$ is unknown.  Recent work suggests that it must 
vary steeply with redshift in order to bring observations of galaxies,
the Lyman-$\alpha$ forest, and the CMB optical depth to Thomson scattering
into agreement~\citep{ino06,fin11,haa12,kuh12}.  We adopt a strongly
redshift-dependent $\fesc$ in our simulations:
\begin{eqnarray}
\fesc = \left\{ 
\begin{array}{lc}
  \fescfive \left(\frac{1+z}{6}\right)^\kappa & z < 10 \\
  1.0 & z \geq 10
\end{array} \right.
\end{eqnarray}
Here, the normalization $\fescfive$ sets the escape fraction at $z=5$ 
and $\kappa$ controls
how strongly $\fesc$ varies with redshift.  For simulations with outflows,
we use $\fescfive=0.0519$ and $\kappa=4.8$.  The normalization is consistent 
with observations of Lyman break galaxies~\citep{nes11} at $z\sim3$, while the 
power-law index lies within 
the range (1--6) that~\citet{kuh12} indicate is required by observations
if the UV luminosity function of galaxies extends to absolute magnitudes 
of -15 (as required by observations of gamma-ray bursts;~\citealt{tre12}).
With this parametrization, the effective optical depth to Lyman-$\alpha$
absorption at $z=5$ is $\ta=3.1$, marginally consistent with the observed 
range of 2--3~\citep{fan06}.  Meanwhile, the predicted ionising emissivities
are (4.6, 7.0) $\times10^{50}$s$^{-1}$Mpc$^{-3}$ at $z=$(5, 6).  At $z=5$,
this is consistent with the observed range of $4.3\pm2.6$~\citep{kuh12}. 
At $z=6$, however,  it overshoots the observed limit $<2.6$, suggesting 
that the predicted emissivity should strengthen rather than declining 
toward the end of reionisation.  Note that our model is not unique in 
failing to reproduce the observed emissivities at $z=5$ and $z=6$ 
simultaneously; this evolution is evidently quite strong compared to 
expectations from models (see, for example, Figure 10 of~\citealt{kuh12} 
or Figures 8 and 15 of~\citealt{haa12}).  The resulting reionisation 
history corresponds to an optical depth to Thomson scattering of 0.071, 
only slightly below the observed $1\sigma$ confidence intervals of 
$\tes=0.088\pm0.015$~\citep{kom11}.

Simulations without outflows require a steeper power-law index and a lower 
normalization because they overproduce the observed galaxy abundance~\citep{dav06}.  
We adopt $\fescfive=0.0126$ and $\kappa=7.21$.  With this dependence, 
they predict emissivities of 5.4 and 7.8 $\times10^{50}$s$^{-1}$Mpc$^{-3}$ at $z=5$ 
and 6, respectively, once again indicating good agreement at $z=5$ 
but incorrect evolution to higher redshift.  The predicted Lyman-$\alpha$ 
optical depth at $z=5$ is 2.2, well within the observed range.  
The predicted Thomson scattering optical depth is 0.074, within 
the observed $1\sigma$ confidence intervals.

The physical interpretation of an evolving $\fesc$ could take many
forms.  For example, direct measurements of ionising continuum flux
from galaxies at $z\approx3$ suggest that current stellar 
population models underestimate the ionising luminosity of 
low-metallicity Population II stars~\citep{nes11}.  If confirmed, 
this observation could imply an evolutionary trend to higher ionising 
luminosity at high redshift that mimics an evolving $\fesc$. 
Alternatively, observations indicate that galaxies grow bluer 
and by inference less dusty beyond $z=4$~\citep{fin10}; this could
drive up $\fesc$ if dust dominates the absorbing column 
(although~\citealt{gne08} argue that $\fesc$ is dominated by 
the ISM's geometry rather than its dust content).  Finally,
galactic outflows can fill a halo with gas, potentially 
creating a significant optical depth to ionising photons.  
If these screens grow weaker with increasing 
lookback time, then the fraction of ionising photons that travel 
unimpeded from the edge of a galaxy's ISM to the virial radius
could grow accordingly; this would mimic an evolving $\fesc$.  
We defer detailed investigation of these possibilities to future 
work; for the present, we simply choose a parametrization that 
will improve the realism of our simulated density field.
We discuss drawbacks to this approach in
Section~\ref{sec:disc}.  

We assume that each photoionisation deposits 4.08 eV of latent 
heat into the IGM.  This heats newly-reionised gas to 
$\approx15,000$ K, which lies within the range that is expected 
for a Population II stellar spectrum.  Each radiation 
transport cell's opacity is given by its volume-averaged 
neutral hydrogen density (counting only the gas that is not
self-shielded; see below) multiplied by the cross-section 
for ionisation at an energy of 17.68 eV.

\subsection{Self-shielding Within Overdense Systems}

Two central challenges in using numerical models to compute the
IGM recombination rate involve deciding which gas belongs to the
IGM and ensuring that overdense gas is allowed to self-shield
against the EUVB.

The need to define the IGM in a physically-motivated way is
illustrated by the results of the pathbreaking radiation
hydrodynamic simulations of~\citet{gne97} and~\citet{gne00}.
These simulations faithfully modelled the growth of ionised
regions as well as the hydrodynamic response of the heated gas.
However, the volume-averaged recombination
rates were computed over all the gas in the simulation, leading to
recombination rates that were much larger than what is plausible for 
the moderately overdense ($\rho/\langle \rho \rangle<100$) gas that 
is conventionally associated with the IGM.  It also led to a 
strong dependence of the recombination rate (and the reionisation 
photon budget) on the simulation dynamic range, with higher-resolution 
simulations absorbing more photons~\citep{gne00}.

Improvement results from using simple density cuts to isolate
low-density gas in post-processing from precomputed simulations.  
This approach has been applied both to N-body 
simulations~\citep{ili07,rai11} and to cosmological hydrodynamic 
simulations~\citep{paw09a,shu12}.  It reduces $C$ as expected,
but the results remain dependent on the precise cuts used to 
define the IGM.  In the case of hydrodynamic simulations, additional 
uncertainty results from assuming a homogeneous EUVB, which may 
pressurize regions that are in reality dense enough to self-shield
and does not account for the spatial inhomogeneity of reionisation.  
Both of these effects could modify the IGM's density and temperature 
fields~\citep{fur09}, which in turn impact the recombination rate.

In this work, we combine these two approaches by using radiation
hydrodynamic simulations of the reionisation epoch that incorporate
a physically-motivated treatment for self-shielding within overdense
regions.  Effectively, we incorporate directly into our simulations
the definition that \emph{the IGM consists only of that gas whose 
ionisation state is not determined by local sources}; denser systems 
are allowed to remain neutral.  By isolating overdense systems from 
the EUVB, we (1) improve the accuracy of our predicted density and 
temperature fields; (2) prevent our simulations from overestimating 
the reionisation photon budget; and (3) obviate the need to choose 
a self-shielding threshold overdensity (such as 50 or 100) when 
deriving the IGM recombination rate in post-processing: the 
threshold evolves on-the-fly in a way that follows the (local) 
gas temperature and (global) EUVB amplitude.

We model the self-shielding of overdense regions using a subgrid
recipe because our radiation transport solver does not resolve
them spatially even though our hydrodynamic solver does.  For 
example, Lyman limit systems (with neutral column densities of 
$\sim10^{17}$cm$^{-2}$) have a characteristic size of 10 physical 
kpc~\citep{sch01,mcq11} and must be treated with 5--10 resolution 
elements~\citep{gne06}.  In our 
higher-resolution simulation (r6n256wWwRT32), the gravitational softening 
length is $0.1$ kpc (Plummer equivalent) while the radiation transport solver achieves 
a resolution of 38 kpc (where both numbers are in physical units 
at $z=6$).  Hence while our hydrodynamic solver certainly resolves
Lyman limit systems, our radiation solver's resolution is a factor 
of $\approx20$ too coarse.  

We overcome this limitation by following 
the argument of~\citet{sch01}.  This work 
showed that, if Lyman limit systems are in hydrostatic equilibrium, 
then they can be identified directly with gas above a critical 
overdensity~\citep[see also][]{mcq11}.  The threshold, $\dlls$, 
is given by:
\begin{equation}\label{eqn:dlls_typ}
\dlls = 25 
\left(\frac{T}{10^4 \rm{K}}\right)^{0.3} 
\left(\frac{1+z}{7}\right)^{-3}
\left(\frac{\Ghi}{2\times10^{-13}}\right)^\frac{2}{3} 
\end{equation}
at the Lyman limit.  Here we have assumed that hydrogen is fully 
ionised and helium is singly-ionised, as expected for soft 
ionising sources~\citep{cia12,fri12}.  $\Ghi$ represents the volume-averaged 
EUVB and does not include the influence of local sources.
This estimate is within a factor of two of what others have 
found~\citep[for example,][]{bol07}.  We do not expect slight
differences owing, for example, to our assumed recombination 
rate to affect our results.  We enforce $\dlls\geq2$ in order to 
preserve self-shielding at $z\gg10$; this floor is not expected to 
affect our results.  We use this approximation to model subgrid 
self-shielding as follows: For each gas particle whose density lies 
below the physical threshold for star formation (0.13 cm$^{-3}$), 
we compute $\dlls$ using the local gas temperature, the case-A 
recombination rates, and the global mean ionisation rate per 
hydrogen $\Ghi$.  We then assume that the optical depth to the 
diffuse background varies as a power-law of the overdensity
\begin{eqnarray}\label{eqn:tau_delta}
\tau_\Gamma = \left\{ 
\begin{array}{lc}
  0 & \Delta < \dlls \\
  \tau_0\left(\frac{\Delta}{\dlls}\right)^b & \Delta \geq \dlls
\end{array} \right.
\end{eqnarray}
In the optically-thin limit, $\tau_0 = 1$ and $b=1.5$~\citep{sch01}.
In reality, $\tau_\Gamma$ increases more rapidly with density because 
the neutral 
fraction in the partially shielded outskirts of an absorber is larger 
than in the optically-thin case.  We have experimented with both 
$b=1.5$ and $b=3$ and found little difference in practice.  We adopt 
the fiducial choices $\tau_0 = 1.0$ and $b = 3$ in this
work.

We implement Equations~\ref{eqn:dlls_typ}--\ref{eqn:tau_delta} directly
into our simulations so that overdense gas is shielded from the
EUVB on-the-fly.  Our use of different discretizations for the 
radiation and hydrodynamic solvers requires that we do so in two
ways.  When updating the ionisation state of a given particle, we
attenuate the radiation field of its host cell by the factor 
$e^{-\tau_\Gamma}$, where $\tau_\Gamma$ is evaluated using the global
EUVB and the particle's temperature.  Conversely, when assembling the 
opacity field on the radiation solver's grid, we reduce each particle's
contribution to its host cell's opacity by the same factor.  We include
the opacity of each gas particle's self-shielded region by treating it 
as an opaque sphere with volume equal to $f\xhi(1-e^{-\tau_\Gamma})\Delta V$,
where $\Delta V$ is the ratio of the gas particle's mass to its 
density, $\xhi$ is its neutral fraction, and $f$ is a parameter 
that is tuned via high-resolution calculations to $1/8$.  For example, 
if $e^{-\tau_\Gamma}=0.5$ for a particle, then its ionisation rate is 
50\% of that of its host cell's, and only 50\% of its neutral 
hydrogens contribute to the cell's opacity.  We also augment the cell's 
opacity by the amount
$4.84(0.5\xhi\Delta V)^{-2/3}/A_{\rm{cell}}$, where $A_{\rm{cell}}$ 
is the cell's area.  Note that the opacity from 
partially-neutral gas particles whose overdensity falls below the 
self-shielding threshold is included directly in the radiation 
transport solver as in our previous work.

Intuitively, this treatment divides self-shielded regions into an optically 
thin skin and an optically-thick core.  The overdensity range over which
a region transitions from optically-thin to optically-thick follows 
from Equation~\ref{eqn:tau_delta}.  The opacity owing to the 
optically-thin region is distributed uniformly throughout the 
radiation transport cell while the core is treated as a photon sink.
This approximation preserves the ability of dense gas 
to remain self-shielded within a coarse radiative transfer grid.
It will enable radiative hydrodynamic simulations to subtend 
cosmological volumes even as they treat the radiation field's small 
scale structure with sufficient detail to model observables such
as the abundance of low-ionisation metal absorbers and the 
post-reionisation IGM opacity.

In reality, regions that are dense enough for star formation to occur 
($\Delta>1000$) contain a mixture of ionised and neutral
gas owing to the local radiation field from massive stars.
They remain entirely neutral in our simulations owing to our self-shielding
prescription.   Accounting for their recombinations would lead to much 
larger clumping factors~\citep{gne97,gne00}.  However, the corresponding
absorptions should be identified with an ionising escape fraction $\fesc$ 
that is less than unity rather than an increased clumping factor 
because they are not dominated by the EUVB.  We encapsulate their
impact via an assumed $\fesc\leq1$ that is tuned using observational 
insight (see above) and defer the challenging task of modeling $\fesc$ 
directly to future work.

\begin{figure}
\centerline{
\setlength{\epsfxsize}{0.5\textwidth}
\centerline{\epsfbox{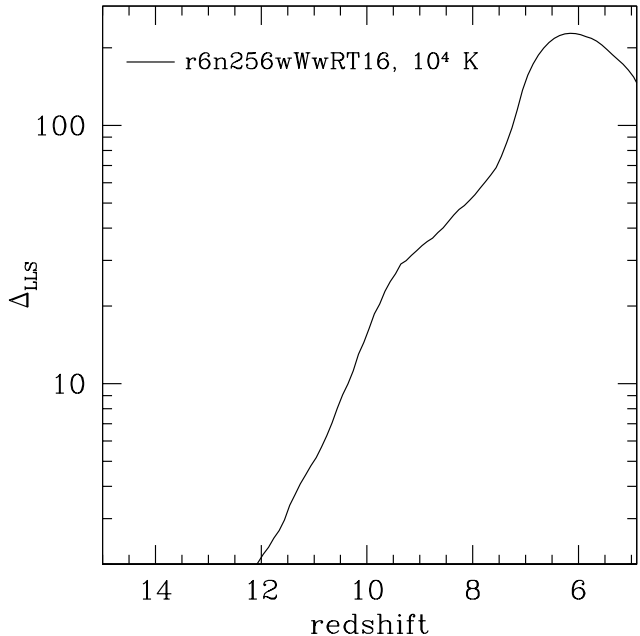}}
}
\caption{The threshold overdensity for self-shielding $\dlls$
as a function of redshift within our fiducial simulation at a 
characteristic temperature of $10^4$K.  $\dlls$ peaks at over
200 during the heyday of reionisation before shrinking to 
roughly 150 by $z=5$ owing to declining escape fractions.
}
\label{fig:dlls}
\end{figure}

Figure~\ref{fig:dlls} shows how $\dlls$ evolves in our fiducial
simulation for a characteristic temperature of $10^4$K.  Note
that this figure is illustrative; in reality our simulations 
compute this threshold on-the-fly using the local temperature.  
$\dlls$ grows from values near unity at the onset of reionisation 
to values near 200 by the time that reionisation
has completed.  Gas that is denser than this threshold sees an
attenuated EUVB, suppressing both the reionisation photon budget
and the volume-averaged recombination rate.  

In detail, the
values in Figure~\ref{fig:dlls} may be somewhat overestimated
(see, for example,~\citealt{mcq11}).  This occurs because our
monochromatic radiation transport solver uses a smaller 
photoionisation cross section $\sigma$ than the value at the 
Lyman limit $\sigma_{\rm{LL}}$ in order to match the assumed 
photoheating rate of 4.08 eV per photoionisation.  In particular,
$\dlls$ varies with the photoionisation cross section $\sigma$ 
as $(\sigma/\sigma_{\rm{LL}})^{-2/3}$~\citep{sch01}, hence our 
simulations may overestimate $\dlls$ by a factor of $\approx1.7$.  
This is roughly the difference between our value at $z=6$ and 
that of~\citet{mcq11}.  The overall impact of this limitation on 
our inferred value of $C$ is weak; we will quantify it in 
Section~\ref{sec:disc}.

Table~\ref{table:sims} introduces our simulation suite.  The simulation
names encode the physics and numerical resolutions.  For example,
the r6n256wWwRT16 simulates a $6\hmpc$ volume (r6) with $2\times256^3$
particles (n256), includes outflows (wW) and evolves the EUVB on
a Cartesian grid with $16^3$ cells (wRT16).  The first four simulations
explore the sensitivity of gas clumping to photoheating
and outflows.  The next three explore convergence with respect to our
radiation and gas solvers as well as the cosmological volume.  The
final simulation assumes an optically-thin EUVB for comparison with
previous work.  Throughout this work, we will use the
r6n256wWwRT16 simulation as our fiducial case.

Our simulations allow us to explore how the IGM's temperature impacts 
its recombination rate within the context of a model that treats 
photoheating from Population II stars self-consistently 
(Section~\ref{ssec:feedbackAll}).  As a check on whether our predictions
are realistic, we compare the temperature at mean
density to recent observations.~~\citet{bol12} used measurements
of the Doppler widths of Ly$\alpha$ absorption lines along the line 
of sight to seven quasars to infer that the IGM temperature at mean 
density is $\log(T)=3.85\pm0.08$ at $z=6.08\pm0.33$ (their Table 3, 
fiducial model).  Our fiducial simulation predicts 
$\log(T,z=6)=3.93\pm0.10$.  Here, we report the median $\log(T)$ over 
particles with overdensity $-0.2<\log(\Delta)<0.2$; switching 
from the median to the mean changes results by $\sim0.1\%$
because the scatter in the temperature distribution is not large.
This comparison is of course incomplete because it does not account 
for a variety of observational systematics.  Nonetheless, the fact 
that the simulation reproduces the observed value within the errors 
indicates that the predicted temperature fields are plausible. 

\section{What is gas clumping?} \label{sec:defs}
The motivation for defining clumping factors comes from a desire to take the
volume-weighted mean of the ionisation rate equations.  For example, the
rate of change of the neutral hydrogen abundance $\nhi$ is given by
\begin{equation} \label{eqn:dnHIdt}
\frac{d\nhi}{dt} = \Ghi \nhi + k_2 \nhii \nem - k_1 \nhi \nem,
\end{equation}
where $\Ghi$ is the photoionisation rate per hydrogen atom, $k_1$ is the
collisional ionisation rate, and $k_2$ is the (case A) recombination rate.  
Each term
on the right side is nonlinear, hence, in principle, it is not possible to compute their
spatial averages unless the cross-correlation of the abundances of $\nhi$ and
$\nhii$ with each other as well as with the radiation and temperature fields
are known (since $k_1$ and $k_2$ are temperature-dependent).  It is convenient
to encapsulate this subgrid information with clumping factors.  For 
example, the spatially averaged recombination rate is approximated as

\begin{equation} \label{eqn:Cdef}
\langle k_2 \nhii \nem\rangle \equiv \cnenptb\langle\nhii\rangle\langle\nem\rangle\langle k_2 \rangle,
\end{equation}
where angle-brackets indicate averages over the entire simulation volume. 
Similar clumping factors could be defined for each of the terms on the 
right-hand side of Equation~\ref{eqn:dnHIdt}.  In practice, however, analytical
calculations bypass the need for the first term's clumping factor by
assuming that the volume-averaged $\Ghi$ equals the ionising emissivity 
divided by the mean hydrogen abundance.  Meanwhile, the third term can
be neglected because $\Ghi \gg k_1\nem$ except within shocks.  This means
that, along with the ionising escape fraction, $\cnenptb$ is one of the major 
uncertainties that hamper efforts to connect the observed abundance of 
ionising sources (such as Lyman Break galaxies) with the ionisation state 
of the IGM (see~\citealt{koh07} for an expanded discussion 
of the other clumping factors).  In Section~\ref{sec:photCount}, we will 
show that the numerical 
results may be reproduced by an analytical model that considers only the 
recombination clumping factor even though the full numerical calculation 
accounts for all three terms.

\begin{figure}
\centerline{
\setlength{\epsfxsize}{0.5\textwidth}
\centerline{\epsfbox{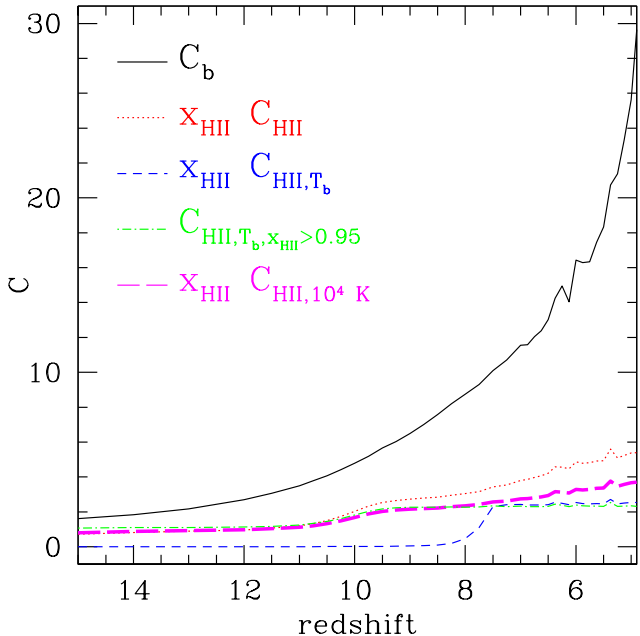}}
}
\caption{The evolution of four different clumping factors in our
fiducial simulation volume.  The clumping factors of all gas whose
density exceeds the star formation threshold reaches 15 by $z=6$.
Accounting for the fact that overdense gas is self-shielded and that
ionised gas is somewhat warmer than $10^4$K leads to much lower clumping
factors (see text for discussion).
}
\label{fig:clumpdefs}
\end{figure}

Equation~\ref{eqn:Cdef} shows that the clumping factor depends on
the density, ionisation, and temperature fields.  Partial clumping
factors are often used when one or more of these fields is 
unavailable.  In Figure~\ref{fig:clumpdefs}, we show how the choice 
of definition impacts the clumping factor's value in our fiducial 
simulation.  The black solid curve shows the 
evolution of the clumping factor averaged over all gas whose density
lies below the adopted threshold for star formation (0.13 cm$^{-3}$),
$\cb\equiv\langle\rho^2\rangle/\langle\rho\rangle^2$.  It increases 
monotonically as overdense regions collapse, with a brief enhancement around
the reionisation redshift.  The enhancement occurs when the EUVB grows strong
enough to photoevaporate gas from photosensitive 
halos~\citep{fin11}, temporarily boosting the mass of gas whose
density lies just below the threshold for star formation.  The 
clumping factor of all gas (including star forming gas) decreases at 
this epoch as expected, but the clumping factor of gas that lies 
outside of galaxies---that is, $\cb$---increases until the 
newly-pressurized gas has expanded into hydrostatic equilibrium.

The clumping factor within ionised regions---which is the important 
quantity for reionisation---is expected to be different from $\cb$
because reionisation begins in overdense regions and progresses
into voids~\citep{fur04,mcq07} in such a way that the last regions to 
remain neutral are overdense~\citep{mir00}.  Furthermore, gas that is 
self-shielded does not contribute to recombinations.
To explore these factors, we compute the clumping factor of electrons 
and protons 
\begin{equation}\label{eqn:chii}
\cnenp\equiv \langle\nem\nhii\rangle/\langle\nem\rangle\langle\nhii\rangle.
\end{equation}

$\cnenp$ can be computed in either of two ways.  One way is to compute the
averages over all ionised gas.  Unfortunately, 
this introduces uncertainty owing to the choice of ionisation threshold.
For example, averaging only over gas particles that are more than 50\% 
ionised gives a slightly different answer than averaging only over 
particles that are more than 95\% ionised because the simulated IGM is 
not a strictly two-phase medium.  The alternative is to take averages 
over the entire simulation volume and then weight by the ionised volume 
fraction $\xhiiv$.  This can be understood by recalling that, in a 
partially-ionised universe with a uniform density, $\cnenp=1/\xhiiv$.  
Therefore, $\xhiiv\cnenp$ indicates the recombination rate in units of 
the recombination rate of a homogeneous universe that has the same
ionisation state, which is the goal of the clumping factor. 

We show this quantity with a red dotted line.  
The tendency for overdense regions 
to remain neutral suppresses $\xhiiv\cnenp$ below $\cb$ at 
all times, with reionisation-epoch values that lie below $\approx6$.
This difference immediately reveals the importance of self-shielding,
which creates a boundary between the ionised IGM and the neutral or
locally-ionised
regions near halos.  It can be modeled directly by spatially resolving 
physical scales of $\sim1$ kpc~\citep{gne06}.  However, this incurs 
significant computational expense, limiting the simulation's 
cosmological volume.  Our approach of coupling a somewhat coarse 
grid for the radiation solver with a subgrid prescription for 
self-shielding may open up the possibility of simulating cosmological 
reionisation within large volumes while treating the IGM's thermal 
and ionisation states faithfully.

$\cnenp$ is an improved description of the volume-averaged 
recombination rate over $\cb$, but it remains incomplete because 
the recombination rate $k_2$ also depends on the gas temperature.  
To illustrate this, we use a blue dashed curve to show the 
temperature-corrected clumping factor $\xhiiv \cnenptb$, computed 
following Equation~\ref{eqn:Cdef} (referred to by~\citealt{shu12}
as $C_{\rm{RR}}$).  This curve remains near zero until 
reionisation is well under way because the recombination rate per 
proton is spatially anti-correlated with the abundance of ionised 
gas.  Once $\xhiiv$ exceeds $\approx0.5$, it begins to climb 
because $\langle k_2\rangle$ is no longer suppressed by cold neutral
regions.  However, it does not reach $\cnenp$ owing to the lingering 
presence of self-shielded regions. Following reionisation, 
$\xhiiv \cnenptb$ climbs slowly because the volume-weighted mean 
temperature is supported by the slow photoheating of filaments, 
which suppresses $\langle k_2 \rangle$.  It never exceeds 3, 
reflecting the tendency of photoheating to suppress
recombination rates.

$\xhiiv\cnenptb$ provides a reasonable description of the clumping 
factor within ionised regions once reionisation is well under 
way, but it is less informative at early times ($\xhiiv < 0.1$)
when the high recombination rate in the predominantly cold, neutral
IGM suppresses $\cnenptb$ despite the high clumping factor of 
ionised regions.  We may compute a more informative clumping factor 
by re-evaluating $\cnenptb$ using only those regions (that is, SPH 
particles) whose ionised mass fraction is greater than 0.95.
This clumping factor (green dot-dashed) is somewhat larger at earlier
times because it is not suppressed by the high recombation rates
per proton in regions that have not yet been heated.  
It merges with $\xhiiv\cnenptb$ once 
$\xhiiv\sim1$.  A drawback of $\cnenptb$ is that its value depends
on the threshold ionisation fraction that is used to compute it.

While $\cnenptb$ emerges naturally from a spatial average of 
Equation~\ref{eqn:dnHIdt}, it is difficult to compare with observations 
because the mean IGM temperature (and hence $\langle k_2 \rangle$) 
at $z=6$ remains poorly-constrained; this is of course equally true
even if $\cnenptb$ is computed only over regions whose ionisation fraction
exceeds some threshold.  To resolve these problems, we also compute 
the ``observational temperature-corrected clumping factor of ionised 
gas" 
\begin{equation}\label{eqn:chiiten}
\cnenpobs \equiv \frac{\langle \nem \nhii k_2(T)\rangle}
	{\langle \nem \rangle \langle \nhii \rangle k_2(10^4\mathrm{K})}
\end{equation}
in which we replace the volume-averaged recombination rate 
$\langle k_2\rangle$ in the denominator with the recombination rate at 
$10^4$K.  This clumping factor (referred to by~\citealt{gne00} as $C_{\rm{HII}}$)
expresses the mean recombination rate 
without requiring knowledge of the topology of reionisation or the 
temperature of ionised gas.  It is slightly higher than $\cnenptb$ 
following reionisation, reflecting the tendency for gas to cool below 
$10^4$ K once reionisation is complete (that is, 
$\langle k_2 \rangle > k_2(10^4\mathrm{K})$ for $z<6$).  The function
\begin{equation} \label{eqn:cfit}
\xhiiv\cnenpobs = 9.25 - 7.21 \log_{10}(1+z)
\end{equation}
fits its evolution to within 30\% for $z\leq15$ and 10\% for $z\leq 10$.

Note that uncertainties in the recombination rate translate directly 
into uncertainties in the clumping factor.  For example, our adopted
(case A) recombination rate is $\sim20\%$ higher than that of~\citet{hui97} 
in the temperature range $10^3$--$10^5$K.  This means that ionisation 
fronts do not penetrate as far into overdense regions in our simulation
as they would if we assumed the~\citet{hui97} rates, reducing the 
inferred clumping factor of ionised gas by the same ratio.  The
resulting uncertainty does not affect our results and is trivial to 
adjust for: Simply divide our inferred clumping factors by the ratio 
of the alternative recombination coefficient to our 
adopted coefficient at $10^4$ K. 

In summary, Figure~\ref{fig:clumpdefs} indicates that the clumping 
factor's value depends rather strongly on its definition.  Taking 
self-shielding into account suppresses it from $\approx30$
to $\approx6$ at $z=5$ because the most overdense regions are self-shielded
and do not contribute to the IGM recombination rate.  Taking the IGM's 
temperature into account further suppresses clumping to 2--4
because photoionised regions are generally warmer than $10^4$ K.

\section{The Role of Feedback}\label{sec:feedback}

Galactic outflows and photoheating both affect 
gas clumping.  Heating suppresses gas clumping by smoothing density
fluctuations on scales smaller than the local Jeans length~\citep{efs92}.
Outflows increase IGM clumping partly by boosting the amount of 
dense gas that resides within halos but outside of galaxies and 
partly (for a given $\fesc(z)$) by delaying reionisation and heating.  
The latter effect is a straightforward consequence of suppressing 
star formation.  The former is expected based on our 
previous finding that outflows boost the number of ionising photons 
required to achieve reionisation~\citep{fin11} although it may in
reality be weaker if ejected gas self-shields.  In this section, we 
quantify how these processes modulate gas clumping.

\subsection{Averaging over the IGM} \label{ssec:feedbackAll}
\begin{figure}
\centerline{
\setlength{\epsfxsize}{0.5\textwidth}
\centerline{\epsfbox{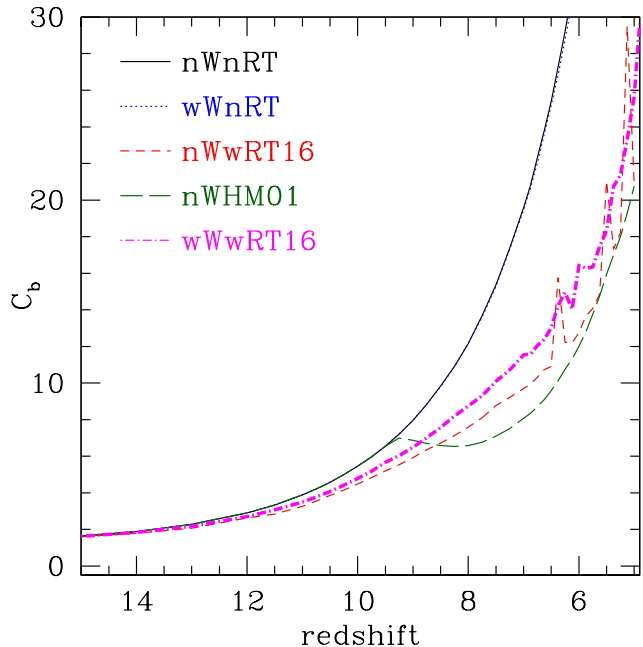}}
}
\caption{The impact of photoheating and outflows on the
overall IGM baryonic clumping factor $\cb$.  All simulations subtend 
a $6\hmpc$ volume discretized with $2\times256^3$ particles (r6n256);
see Table~\ref{table:sims}.  Outflows have negligible impact on $\cb$ 
in the absence of an EUVB, but when acting in concert with an EUVB 
their impact is stronger than uncertainties owing to limitations in 
our particular treatment of the EUVB.
}
\label{fig:cb_feedback}
\end{figure}

We begin by discussing the separate impacts of outflows and
photoheating on the IGM clumping factor.  In order to 
facilitate this discussion, we revert to the definition $\cb$, 
which uses all gas that is too diffuse to form stars 
(Section~\ref{sec:sims}) without reference to
its ionisation state or temperature.  While this definition
includes gas that is dense enough to self-shield and is therefore
an imperfect estimate of the IGM recombination rate, it allows us 
to compare simulations with and without an EUVB.

We use solid black and dotted blue curves in 
Figure~\ref{fig:cb_feedback} to show how $\cb$ evolves in simulations
without and with outflows in the absence of an EUVB.  They are
nearly coincident, indicating that outflows affect the gas density 
distribution only weakly if they do not couple to an EUVB.  Adopting
a spatially-homogeneous~\citet[][hereafter HM01]{haa01} EUVB dramatically 
suppresses the clumping factor following $z=9$ (long-dashed green;
see also~\citealt{paw09a}).  

Our self-consistent simulation omitting outflows (r6n256nWwRT16, short-dashed 
red) predicts a reionisation history in which the neutral hydrogen fraction
drops to 50\% at $z=9.2$, quite similar to the redshift at which the HM01 
EUVB reionises the Universe ($z=9$).  Prior to $z=5$, its $\cb$ exceeds 
the HM01 case owing to its more extended reionisation history.  By $z=5$, 
its $\cb$ converges to the HM01 case although it shows several spikes 
that are likely associated with photoevaporation of star-forming gas 
from low-mass ($M_h < 10^9\msun$) haloes in regions that are just being 
reionised.  This suggests that $\cb$ is weakly sensitive to the 
reionisation history at early times but converges to expectations 
from calculations using a spatially-uniform EUVB fairly rapidly 
(probably on the sound-crossing time of the filaments).

Turning to our fiducial simulation (magenta dot-dashed), which does include
outflows, we find that its $\cb$ is higher than the no-wind case at all
redshifts.  This model has significantly less star formation than the
no-wind model, but it also assumes a higher ionising escape fraction in
order to compensate (Section~\ref{ssec:rt}).  Combining these factors leads
to a reionisation history in which the neutral fraction drops below 50\%
at $z=8.9$, only slightly later than the other models.  Despite
their similar reionisation histories, the wind model's $\cb$ is roughly 
30\% higher at $z=5$.  Hence although the reionisation histories of the 
wind and no-wind models are not exactly the same, 
Figure~\ref{fig:cb_feedback} confirms that outflows enhance $\cb$ by 
boosting the amount of gas that is overdense but not star-forming.

We draw several conclusions from Figure~\ref{fig:cb_feedback}.  First,
outflows do not affect $\cb$ unless they are heated by an EUVB.  Intuitively,
outflows consist largely of material entrained from their host galaxies'
interstellar media, hence they remain relatively cold and dense unless they
are heated by an EUVB.  Second, an EUVB heats and evaporates gas out of 
shallow potential wells such as filaments and minihalos, 
suppressing $\cb$.  Finally, a better understanding of the nature of 
galactic outflows is {\it at least} as important to our understanding 
of the IGM density structure as the nature of the EUVB.  In 
Section~\ref{ssec:feedbackRhomax}, we will ask whether this
uncertainty impacts the recombination rate of diffuse gas.

\subsection{The recombination rate as a function of density} \label{ssec:feedbackRhomax}
\begin{figure}
\centerline{
\setlength{\epsfxsize}{0.5\textwidth}
\centerline{\epsfbox{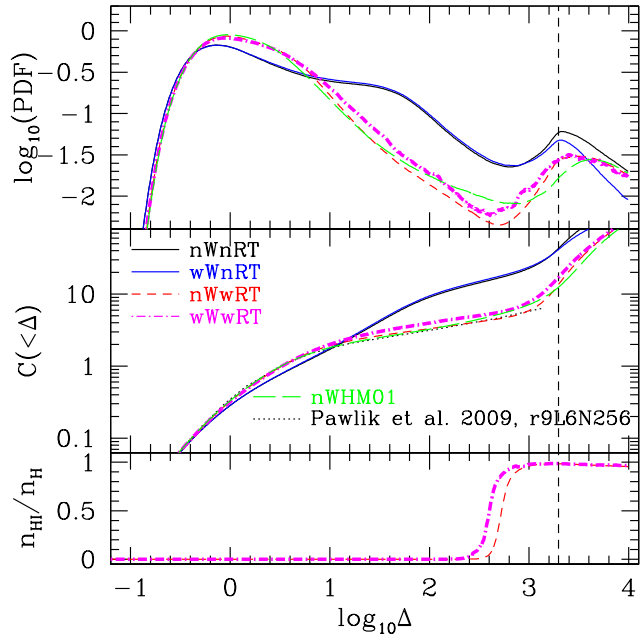}}
}
\caption{(top)
The gas density PDF in units of mass-weighted probability
per unit $\log_{10}(\Delta)$ for six different simulations as indicated by
the legend in the middle panel at $z=6$; all simulations subtend volumes 
with side 
length $6\hmpc$ and resolve halos of total mass $1.4\times10^8\msun$.  
The vertical dashed line indicates the threshold density for star and 
wind formation (0.13cm$^{-3}$).  Photoionisation heating pressurizes 
gas with densities $\Delta>10$, and outflows transfer gas from 
galaxies into the IGM. 
(middle)
The cumulative baryonic clumping factor as a function of maximum
overdensity compared to the fiducial simulation of~\citet{paw09a} 
(black dotted).  An EUVB suppresses the clumping of all gas, 
with minor differences owing to the details of outflows and 
reionisation. 
(bottom) 
The neutral hydrogen fraction as a function of overdensity.  
Gas above a threshold overdensity (which is related to $\dlls$) is 
neutral owing to our self-shielding model.
}
\label{fig:pdf}
\end{figure}

Figure~\ref{fig:cb_feedback} suggests that outflows and photoheating
modulate the IGM's density structure, but their impact on the IGM
recombination rate remains unclear because $\cb$ averages over gas
that is in reality self-shielded.  In order to understand how feedback
processes modulate the abundance of gas at different densities, we 
show the mass-weighted 
probability density function of gas (PDF) at $z=6$ as a function 
of overdensity $\Delta\equiv\rho/\langle\rho\rangle$ in the top 
panel of Figure~\ref{fig:pdf}.  Comparing the blue and black 
solid curves to the lower curves shows how photoheating 
smooths over filaments and removes gas from low-mass haloes 
(see also~\citealt{paw09a}).  At densities higher than the 
adopted threshold for star formation (vertical dashed line), the
PDF is depleted by star formation and outflows.

Above $\Delta\sim240$, both reionisation simulations (short-dashed
red and dot-dashed magenta) show an increasing gas abundance because 
the EUVB is attenuated in these regions owing to self-shielding.
Outflows modestly increase the gas abundance throughout the region
$10 < \Delta < 1000$.  The long-dashed green curve shows the PDF 
from the r6n256nWHM01 simulation, which omits outflows and assumes 
a spatially-homogeneous EUVB (with no self-shielding).  Its gas 
abundance is suppressed with respect to the self-shielding 
calculations in regions where $\Delta>1000$ and enhanced near 
$\Delta\sim200$ because the EUVB evaporates dense gas that should 
in reality be shelf-shielded.  

Differences in the gas density PDFs translate into different amounts
of gas clumping.  In the middle panel of Figure~\ref{fig:pdf}, we show 
the cumulative clumping factor as a function of the threshold 
overdensity $C(<\dtr)$, defined following~\citet{paw09a}:
\begin{equation}\label{eqn:dtr}
C(<\dtr) \equiv \frac{\int_{0}^{\dtr} d\Delta \Delta^2 \mathcal{P}(\Delta)}{\int_{0}^{\dtr} d\Delta \mathcal{P}(\Delta)}
\end{equation}
Here, $\mathcal{P}(\Delta)$ is the volume-weighted probability 
distribution function of gas density.  Note that each curve's value at 
the star formation threshold equals the corresponding simulation's 
$\cb$ at $z=6$ in Figure~\ref{fig:cb_feedback}.  
The solid blue and black curves 
confirm that outflows have little impact on the clumping factor without 
an EUVB.  Adopting a spatially-homogeneous EUVB (green long-dashed) 
reproduces the~\citet{paw09a} result (black dotted) as expected, with 
slight differences likely owing to their addition of extra heating 
around the reionisation redshift.  Our radiation-hydrodynamic 
simulations again predict very similar clumping factors.  Outflows
boost clumping as expected from the top panel, but the change is slight.  
In particular, we confirm the finding by~\citet{paw09a} that outflows change 
$C(<\dtr)$ by $<10\%$ when averaged over gas with $\Delta<100$.  This is 
somewhat nontrivial given that our outflow prescription ejects roughly twice 
as much gas per unit stellar mass formed as theirs at this redshift.  
Overall, the impact of changing the EUVB and the outflow model on 
$C(<\dtr)$ is quite modest.  The fact that the gas density PDF at low 
densities ($\Delta < 200$) is relatively robust to these factors
indicates that the recombination rate in the diffuse IGM---which is the
important quantity for understanding the progress of reionisation---is
only weakly sensitive to the uncertain details of star formation and
feedback.  At higher densities, the impact of feedback and self-shielding
is much stronger (top panel), but recombinations in these regions do not 
count toward the IGM recombination rate as their ionisation state is 
dominated by local sources.

The ability of outflows to boost the abundance of neutral gas at moderate 
overdensities ($\Delta > 240$) rather than only at the outskirts of 
galaxies ($\Delta > 1000$) suggests that they could function as 
an additional population of Lyman limit systems residing near or 
within halos.  If so, then they would have to be self-shielded against 
the EUVB.  We show in the bottom panel the neutral hydrogen fraction as a 
function of $\Delta$ for our fiducial simulation.  The neutral fraction 
jumps from 0 to 1 at $\Delta\approx240$, indicating that outflows could 
indeed be partially self-shielded.  This possibility is consistent 
with direct observations of cold gas 
in galactic outflows~\citep[for example,][]{mar05,wei09} as well as
recent theoretical results~\citep{van12}.

In summary, an EUVB significantly suppresses IGM clumping by 
pressurizing moderately overdense regions while outflows modestly 
boost the recombination rate by 
returning gas from galaxies into the IGM.  Meanwhile, self-shielding 
hides the most overdense gas from the EUVB.  This boosts the mass
fraction of neutral gas while suppressing the clumping factor in 
ionised regions.  Outflows are at least partially self-shielded,
suggesting that they could contribute to the IGM opacity.

\section{Application I: Photon-Counting}\label{sec:photCount}
Analytical models of reionisation use clumping factors to account 
for IGM inhomogeneities~\citep{mad99}, prompting the need for
theoretical insight into the nature of clumping.  We have discussed the
various effects that modulate the value of the clumping factor that we
derive from our simulations.  In order to close the loop, we now ask 
whether these clumping factors do indeed constitute a good model for
the IGM recombination rate during reionisation.  We do this by 
using an analytical model to explore how well several different 
clumping factors fare in reproducing the reionisation history of our 
r6n256wWwRT16 simulation.  We will show that clumping factors perform 
reasonably well, and that they fare better if they incorporate information 
from the simulated temperature and ionisation fields.

We use as our model Equation 20 of~\citet{mad99} (this can also be
derived by taking the spatial average of Equation~\ref{eqn:dnHIdt} and
normalizing by the mean hydrogen number density):
\begin{equation} \label{eqn:dnHIdtmean}
\frac{d\xhiv}{dt} = -\dot{n}_\gamma + \mathrm{recombinations}.
\end{equation}
Here, $\xhiv\equiv\langle \nhi/\nh \rangle$ is the volume-averaged 
neutral hydrogen fraction while $\dot{n}_\gamma$ is the ionising 
luminosity per hydrogen atom into the IGM.  We will use the value 
for $\dot{n}_\gamma$ that is predicted
directly by our simulation.  The form of the recombination term varies 
depending on the clumping factor definition.  

There are three caveats to this widely-used formalism.  The first is that
the second term represents a \emph{mass-average} while the third term is 
generally computed in a \emph{volume-averaged} way (see, for example, 
Equation~\ref{eqn:Cdef}).  To see this, recall that the second term is
meant to model the rate of growth of the ionised volume 
fraction~\citep{mad99}.  The rate at which an 
ionised volume $V_I$ grows depends on the ratio of the ionising luminosity 
$\dot{N}_\gamma$ to the gas density at the position of the ionisation front 
$n_H$, $\dot{V}_I = \dot{N}_\gamma/n_H$.  In practice, however,~\citet{mad99} 
substituted the volume-averaged gas density $\langle n_H \rangle$ into the 
denominator.  This approximation is only accurate if the density is
homogeneous or if the sources are widely separated (as in the case of 
helium reionisation;~\citealt{mcq09}), hence it is equivalent to
assuming that the neutral mass and volume fractions are equal, 
$\xhim=\xhiv$.  It is possible to use knowledge of the 
distribution of gas densities and the topology of reionisation (for 
example,~\citealt{mir00}) to compute the $\dot{V}_I$ from the ionising 
emissivity (or, alternatively, to compute the recombination rate from 
the ionised mass fraction), but these steps are omitted in analytical 
calculations.  Owing to this approximation, we will refer only to the 
neutral fraction $\xhi$ rather than to $\xhim$ or $\xhiv$ within the 
context of our analytical model.  The second caveat is that 
Equation~\ref{eqn:dnHIdtmean} omits collisional 
ionisations because they are nontrivial to model 
analytically.  Finally, the third term requires knowledge of the IGM's
density and temperature distributions that is generally encapsulated via 
clumping factors.  All of these simplifications are relaxed in our numerical 
calculations.  In this section, we will use our simulations to ask 
whether they are indeed valid.
\begin{figure}
\centerline{
\setlength{\epsfxsize}{0.5\textwidth}
\centerline{\epsfbox{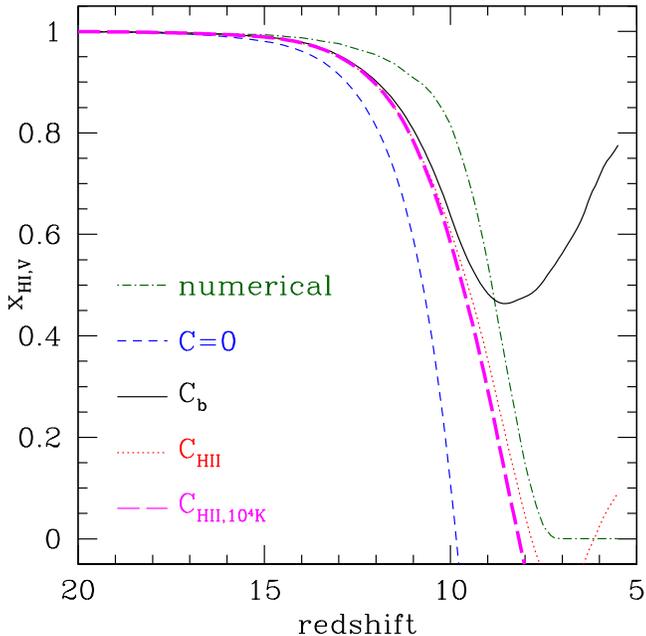}}
}
\caption{The dependence of the reionisation history on the definition of the
clumping factor, as compared to our fiducial radiation hydrodynamic simulations.
In photon-starved reionisation scenarios, a rapidly-growing clumping factor 
can lead to a double-reionisation history although this does not happen with
our preferred definition of the clumping factor.
}
\label{fig:photCount}
\end{figure}

We compare the resulting reionisation histories with our numerical simulation
in Figure~\ref{fig:photCount}.  As a baseline, the short-dashed blue curve 
illustrates the reionisation history if there are no recombinations ($C=0$). 
Comparing it with the dot-dashed green curve indicates that recombinations 
delay the completion of reionisation from $z\approx10$ to $z\approx7.5$.  
The delay is expected given that this simulation requires 3.5 ionising 
photons per hydrogen atom to reach a volume-averaged neutral fraction 
of $\xhi=0.01$.  The other colored 
curves represent three different ways of treating recombinations.  The 
solid black curve uses the clumping factor of all IGM gas and computes 
the recombination term in Equation~\ref{eqn:dnHIdtmean} as:
$\cb k_1(10^4\mathrm{K})\xhii\langle\nh\rangle$, where $\xhii = 1 - \xhi$ 
represents the ionised fraction.  Down to $z=10$, this 
crude treatment is already a significant improvement over the $C=0$
curve.  Below $z=10$, recombinations begin to win over the declining
ionising emissivity, which is in turn driven by the assumed $\fesc(z)$; 
the result is a multimodal reionisation history.  The fact that the 
simulation does not, in fact, yield a multimodal reionisation history 
emphasizes the importance of allowing overdense regions to self-shield 
against the ionising background.  The dotted red and long-dashed 
magenta curves correspond to the definitions $\cnenp$ and $\cnenpobs$ 
from Section~\ref{ssec:feedbackAll}.  In these models, the 
recombination term is 
$C k_1(10^4\mathrm{K}) \xhii^2 \langle\nh\rangle$.  They are each a 
clear improvement over using $\cb$.  Recall that the dotted red curve 
accounts for the order in which regions of different overdensities 
reionise while the long-dashed magenta curve additionally accounts 
for the impact of temperature fluctuations.  Including this information 
accelerates reionisation by suppressing recombinations
(Figure~\ref{fig:clumpdefs}).

In detail, reionisation is more extended in the numerical model even 
when the clumping factor accounts for both the temperature and the 
ionisation fields (long-dashed magenta).  We attribute this to the fact 
that the analytical model effectively assumes an infinite speed of light 
$c$.  We have previously shown that artificially boosting $c$ causes 
reionisation to complete earlier~\citep{fin09b} although the effect 
was slightly weaker ($\Delta z = 0.1$) than what is seen here 
($\Delta z\approx0.9$).  This delay is only a few times larger than 
the light travel time across the photon mean free path at this epoch
($\sim10$ Mpc;~\citealt{fur05}), consistent with the possibility
that light-travel time effects delay the completion of reionisation.

There are two additional effect that extend the final stages of 
reionisation in numerical models.  The first results from the fact 
that the timescale for ionising an absorber varies inversely with its
geometrical cross section.  For example, an EUVB with amplitude $\Gamma$ 
at the frequency corresponding to an absorption cross-section $\sigma$
ionises a spherical absorber of uniform density $n$ and radius $r$ 
on a timescale $\frac{2}{3} n\sigma r/\Gamma$ compared to $1/\Gamma$
in the optically-thin case.  Approximating Lyman limit systems as
uniform spheres in hydrostatic equilibrium at a temperature of 
$10^4 K$, we find that systems with overdensities of 10 and 100 
are ionised by a background $\Gamma=2\times10^{-13}$s$^{-1}$ on
timescales of $\approx$ 8 and 24 Myr at $z=6$, compared to $\approx0.2$
Myr in the optically thin case.  This effect is omitted in analytical
models but present in numerical simulations, although it probably 
delays the completion of reionisation by only $\Delta z\approx 0.1$.

The second delay results from the fact that photons with high energy 
and small absorption cross section take longer to be absorbed than 
photons near the Lyman limit once the ionised volume fraction is large.
This effect is missing from our simulations owing 
to our monochromatic radiation transport solver.  However, it is not 
large: for $z\geq6$, energies less than 54.4 eV, and neutral hydrogen 
fractions greater than 0.01, the delay is less than 20 Myr.  

We may quantify the error in the reconstructed reionisation history using
the optical depth to Thomson scattering $\tes$.  We compute this for the
simulations and the analytical models assuming that helium is 
singly-ionised with the same ionisation fraction as hydrogen down to 
$z=3$~\citep{cia12,fri12}, and doubly-ionised thereafter.  In our 
simulation, the predicted value is 0.071.  Ignoring 
recombinations yields $\tes=0.095$, which confirms that 
galaxies could have dominated reionisation modulo uncertainties regarding
the star formation efficiency of low-mass halos and the IGM recombination 
rate.  Adopting realistic clumping factors measured directly from our 
simulation yields values for $\tes$ that are between 0.062 (solid black) 
and 0.082 (long-dashed magenta).  The discrepancy between the long-dashed
magenta and dot-dashed green curves can be regarded as an estimate of the
uncertainty in analytical reionisation calculations owing to complications
such as light-travel time effects, shadowing, and source clustering; 
it is $\approx10\%$.  In short, analytical models of reionisation 
can reproduce the behavior of more realistic models reasonably well, with 
improvement possible if the adopted recombination rate accounts 
for the inhomogeneous temperature and ionisation fields.  Our analytical 
modeling also confirms that collisional ionisations and the other 
complications that we have mentioned are subdominant.

We note that~\citet{gne00} previously performed an extensive comparison
between his numerical radiation hydrodynamic simulations and analytical models,
finding broad support for the basic assumptions underlying analytical models.  Our
results agree with his.  The principal difference is that, in computing
the volume-averaged recombination rate, their analysis did not distinguish 
between gas within galaxies and the IGM. This led to a much higher value 
for the clumping factor (100--200) than is appropriate for densities 
that are conventionally associated with the IGM (for example, 
$\Delta < 1000$).  Our simulations sidestep this complexity by using
$\dlls$ as a physically-motivated model for the boundary between the IGM 
and the condensed gas whose absorptions are more appropriately 
attributed to $\fesc$.

\section{Application II: The Photon Cost of Reionising Filaments}\label{sec:photCost}
\begin{figure}
\centerline{
\setlength{\epsfxsize}{0.5\textwidth}
\centerline{\epsfbox{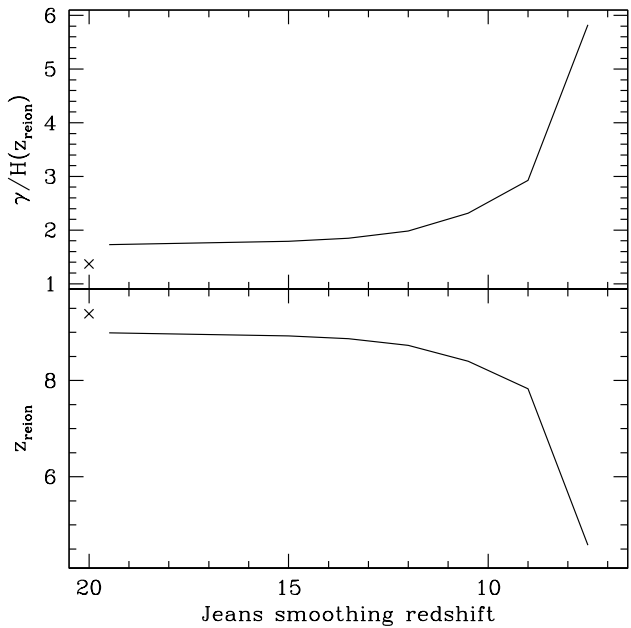}}
}
\caption{(bottom) The dependence of the overlap redshift on heating redshift:
Delaying photoheating from $z=19.5$ to $z=7.5$ delays overlap by $\Delta z=3$.
(top) The reionisation photon budget per hydrogen atom as a function of heating redshift:
Delaying photoheating can more than double the photon cost of reionisation.
The crosses correspond to a homogeneous IGM ($C=1$).
}
\label{fig:photCost}
\end{figure}
While much has been learned about the photon cost of removing 
gas from minihalos~\citep{hai01,bar02,sha04,ili05,cia06}, the ability 
of moderately-overdense gas to delay reionisation has received 
less attention.  This is potentially important for two reasons.  First, 
analytical models may underestimate the reionisation photon budget by deriving 
the clumping factor from hydrodynamic simulations in which the IGM is 
effectively pre-heated.  For example, it is possible to compute the clumping 
factor at high redshift by applying Equation~\ref{eqn:dtr} to the baryonic 
density PDF of~\citet{mir00}~\citep{wyi08,pri10}.  However, that PDF 
derives from simulations in which the diffuse IGM is already 
smoothed by an EUVB.  In reality, there is a photon cost associated 
with smoothing the IGM because the initially cold IGM has a higher 
clumping factor (Figure~\ref{fig:pdf}).  Ignoring this cost may 
cause models to underestimate the photon budget of reionisation.  
Alternatively, it is possible that the IGM is heated before 
reionisation is well under way.  For example, an early X-ray 
background could suppress gas clumping prior to the onset of 
reionisation~\citep{oh03}.  In this case, filaments would absorb fewer
ionising photons than in models (such as ours) in which the primary 
heating source is UV photons.  In this section, we estimate the 
magnitude of this uncertainty by asking how many ionising photons are 
required to reionise filaments, or regions with overdensities of 
$\Delta=$1--50.

Our approach involves decoupling reionisation from heating and 
determining how reionisation changes if the universe is heated at 
different redshifts.  If the universe is heated at high redshift 
(such as $z=19.5$), then the photon cost of reionisation is smaller 
because the universe remains smooth after being heated.  By contrast, 
if the universe is heated at a lower redshift (such as $z=7.5$), then 
reionisation is more expensive because more of its gas has condensed into
filaments by the time that reionisation is under way.  By comparing these 
histories, we gain some insight (albeit not completely self-consistently) into 
the photon price that the universe pays for waiting to ionise its filaments; 
this is probably comparable to the actual photon cost of reionising
moderately-overdense regions.

\citet{paw09a} provide fitting functions for the baryonic clumping factor 
in simulations where an instantaneously imposed spatially-homogeneous 
EUVB is used to heat the IGM at a variety of redshifts.  We have 
combined their fitting functions (choosing their $C_{50}$) with the 
predicted ionising emissivity from our r6n256wWwRT16 simulation 
within our analytical model and evaluated how reionisation
depends on the smoothing redshift.  We show in the bottom 
panel of Figure~\ref{fig:photCost} how the reionisation redshift (where 
$\xhi\rightarrow 0.01$) varies.  Delaying smoothing from $z=19.5$
to $z=7.5$ delays reionisation by $\Delta z \approx 4.5$.  This relatively 
strong dependence owes to the extended reionisation history that is 
enforced by our choice of $\fesc(z)$.  We have verified that extracting
our emissivity history from simulations that assume a constant 
$\fesc=0.5$ yields a smaller delay of $\Delta z = 1.1$ (not shown) 
because the growth rate of collapsed matter (such as stars) is much 
faster than the growth rate of moderately overdense structures 
(such as filaments).

The top panel maps the reionisation redshift into the number of ionising 
photons consumed per hydrogen atom.  If the IGM is 
heated at $z=19.5$, reionisation consumes 1.8 photons per 
hydrogen.  Observations suggest that reionisation was well under way 
by $z=$9--10~\citep{pri10,kuh12}.  Adopting this as the photoheating
redshift yields a total photon cost of 2.5--2.9 photons per hydrogen.  
Subtracting, we find that the universe pays a price of 0.7--1.1 photons 
per hydrogen for waiting to heat its filaments.  Put
differently, analytical models that assume that the universe is smoothed
at a high redshift underestimate the total photon cost of reionisation
by 0.7--1.1 photons per hydrogen atom.

These calculations may underestimate the total photon cost of reionisation 
because they do not account for gas that is condensed into minihalos,
or halos with virial temperatures below $10^4$ K ($\approx10^8\msun$ for
redshifts of 6--10).  Prior to reionisation, much of the gas may have
been condensed into halos with virial masses between $10^6$ and $10^8\msun$.
The clumping factors of~\citet{paw09a} were derived from simulations that
resolved halos at the upper end of this mass range with 100 particles,
hence lower-mass systems are effectively missing.  These halos can 
consume up to 5 photons per hydrogen atom~\citep{sha04}, with the 
implication that the amount of gas clumping could have been much 
higher if the IGM temperature did not exceed 1000 K prior to reionisation.

On the other hand, an early X-ray background could deposit an early 
entropy floor due to the large mean free path of 
X-rays~\citep{oh01,ven01,ric04,mad04}.  Such a background is widely
predicted in many models of reionisation and could arise from X-ray 
binaries, inverse Compton scattering in supernova remnants, or early 
mini-quasars.  This process significantly reduces gas 
clumping~\citep{oh03}, and if it occurs well before the reionisation 
epoch then it could decrease the reionisation photon budget.

In short, while filaments may have absorbed $\sim1$ additional photon 
per hydrogen, the total number of photons required to complete reionisation 
remains uncertain owing to the unknown abundance of small absorbers.
Future work incorporating X-ray heating and a wider dynamic range will 
be required in order to explore these processes.

\section{Discussion} \label{sec:disc}

Could galaxies have produced enough ionising photons to reionise the
Universe? Could they have maintained the IGM's ionisation state at $z=6$?
Two major uncertainties hamper efforts to connect observations of 
galaxies with measurements of the neutral hydrogen fraction in order
to address these questions.  The first is the fraction of ionising 
photons that escape galaxies' ISMs, $\fesc$, and the second is the 
recombination rate of the IGM, which is often parametrized by a 
clumping factor $C$.  The observed population of galaxies can 
maintain an ionised IGM if the ratio $C/\fesc$ lies within the range 
1--10~\citep[][Figure 15]{fin10}.  

We previously showed
that our simulations reproduce the observed UV luminosity function of
galaxies and self-consistently complete reionisation by $z=6$ when we 
assume $\fesc=0.5$~\citep{fin11}.  While encouraging, those simulations 
suffered from two drawbacks.  First, they overproduced the observed
EUVB amplitude at $z\leq6$ and underproduced $\tes$.  Second, the 
radiation transport solver's limited spatial resolution did not shield
moderately-overdense regions from the EUVB, with the result that the
predicted clumping factor of ionised gas was artificially high 
($\approx12$ at $z=6$).  Consequently, completing reionisation 
($\xhiv<0.01$) required up to 5 ionising photons per hydrogen atom.

In this work, we have improved on those calculations by adopting a 
subgrid self-shielding treatment and a redshift-dependent $\fesc$.  
Self-shielding limits the clumping factor and the reionisation photon 
budget by isolating overdense gas from the EUVB.  Meanwhile, an 
evolving $\fesc$ (1) reconciles the model with the observed $\tes$ by
increasing the electron fraction at early times; (2) heats
the IGM at an earlier redshift, which suppresses gas clumping at
later times; and (3) raises the Gunn-Peterson optical depth at $z=5$
into improved agreement with observations.  Together, we find that 
these 
modifications suppress $C$ to values in the range 2.7--3.3 at $z=6$
(the uncertainty comes from possibly overestimating
the overdensity at which gas fully self-shields; see below).
For our adopted $\fesc(z)$, this leads to $C/\fesc\approx$25--30
at $z=6$, slightly too high for currently-observed galaxy 
populations to maintain a completely ionised IGM.  Our 
simulations overcome this barrier by resolving galaxies down 
to absolute magnitudes of $M_{\mathrm{UV}}=-15$, which is 2--3 
magnitudes fainter than current limits for observing galaxies 
directly but consistent with requirements inferred from other 
star formation tracers~\citep{tre12}.

Despite these improvements, there are a number of ways in which we 
can build on this work in the future.  First, we use the global EUVB 
to compute $\dlls$.  Near sources, this underestimates the EUVB, 
leading to an overdensity threshold that is too low.  In voids, it 
overestimates the EUVB, leading to too little 
self-shielding~\citep[see also][]{cro11}.  It would be preferable 
to estimate the EUVB by averaging
over regions that are a few times larger than the expected length scale of 
self-shielding systems in order to incorporate the EUVB's small-scale 
fluctuations into the ionisation field.  This improvement would result in
more-ionised overdensities and more-neutral voids (or in other words a 
more strongly inside-out reionisation topology), but the overall impact
on $C$ is difficult to predict.

Second, our results would benefit from an improved understanding 
of the IGM's temperature $T$.  The IGM recombination rate is 
sensitive to $T$ through the recombination coefficient $k_2$ and 
through the minimum overdensity $\Delta$ at which gas is neutral, which 
in turn is proportional to the self-shielding threshold $\dlls$.  
For temperatures between 
$10^{3.5}$--$10^{5.5}$, $k_2\propto T^{-0.77}$.  Meanwhile,
$C(<\dtr) \propto \Delta^{0.25}$ (Figure~\ref{fig:pdf}) and
$\dlls\propto T^{0.3}$ (Equation~\ref{eqn:dlls_typ}), leading 
to a weak overall dependence
of $C(<\dtr)\propto T^{0.075}$.  Combining these, we find 
that the overall IGM recombination rate varies as $T^{-0.7}$.

Uncertainty in $T$ stems from the unknown latent heat of photoionisation 
and from resolution limitations.  We have effectively assumed 
that reionisation heats gas to $\approx15,000$ K.  While this assumption
yields a mean IGM temperature that is consistent with observations 
(Section~\ref{sec:sims}), the observations are themselves uncertain 
owing to the necessity to correct the measured IGM temperature for 
the influence of nearby quasars~\citep{bol12}.  Assuming a (harder, 
softer) ionising background would lead to a (hotter, colder) 
post-reionisation IGM.  Additionally, the limited spatial 
resolution of our radiation solver may cause our 
simulations to underestimate the post-reionisation gas temperature 
because ionisation fronts are artificially broadened, increasing the 
amount of time during which partially-ionised gas can cool through 
collisional excitation of neutral hydrogen~\citep{mir94}.  The magnitude 
of this effect is likely small compared to the uncertain latent heat of 
photoionisation.  Nevertheless, improved measurements of the IGM 
temperature at $z>5$ would constrain our model further.

Third, our simulations expose somewhat too much overdense gas to the EUVB,
boosting the IGM recombination rate.  For example, the 
threshold overdensity for self-shielding at $10^4$K and $z=6$ should 
be $\dlls=25$ (Equation~\ref{eqn:dlls_typ}).  If the neutral column 
density of self-shielded systems varies strongly with $\Delta$, then 
systems with overdensity more than a few times greater than $\dlls$ should 
be essentially neutral.  For example, the high-resolution calculations
of~\citet{mcq11} indicate that the neutral fraction should increase
rapidly above $\Delta\sim100$ (their Figure 3).  By contrast, the gas 
in our simulations remains ionised until $\Delta\approx200$ 
(Figure~\ref{fig:pdf}), which may be slightly too high.  If so, 
then it owes largely to the fact that our simulations assume that 
all photons have an energy of 1.3 times the threshold energy for 
photoionisation $h\nu_{\rm{LL}}$ in order to photoheat the IGM.  As we 
argued in the discussion of Figure~\ref{fig:dlls}, this assumption may 
artificially boost $\dlls$ by a factor of 1.7.  Invoking once 
again the scaling $C\propto\dlls^{0.25}$ (Figure~\ref{fig:pdf}), we 
find that,
if the amplitude of the ionising background were unchanged but the
radiation field were dominated by photons with energy nearer to 
$h\nu_{\rm{LL}}$, then the clumping factor of ionised gas would be
lower by $\approx12\%$.  Combining this uncertainty with the range of 
values inferred from our fiducial and high-resolution simulations 
(r6wWwRT16 and r6wWwRT32), we conclude that the clumping factor at 
$z=6$ lies between 2.7--3.3.

Finally, our limited cosmological volume forces us to adopt a rather
dramatic evolution for $\fesc(z)$.  We do this because our predicted $\tes$ 
and $\ta$ are not converged with respect to simulation volume or mass 
resolution, as is well known for simulation volumes of this 
size~\citep{gne06,bol09}.  This is not to say that the need for a 
decreasing $\fesc(z)$ is purely a resolution limitation; empirical 
arguments strongly support such a trend and in fact our adopted
relation lies within the observationally-inferred range~\citep{kuh12}.
However, with a larger volume, structure formation
would begin sooner, allowing us to
adopt a weaker redshift-dependence for $\fesc(z)$ without compromising
the agreement between the observed and predicted $\tes$ and $\ta$.  
This would change the predicted IGM temperature-density relationship
as well as the overall gas density distribution.  Both of these changes 
could modify the predicted IGM recombination rate, although the sign
of the effect is difficult to predict.

\section{Summary} \label{sec:sum}
We have used a suite of cosmological radiation hydrodynamic simulations 
to explore the impact of outflows and an EUVB on the recombination rate 
of the reionisation-epoch IGM.  Feedback processes modify 
the IGM's gas density, ionisation, and temperature fields.  We may 
illustrate their impact on the gas density distribution by considering 
the clumping factor 
$C_{100}\equiv\Delta^2 \mathcal{P}(\Delta)$ at redshift $z=6$ and
overdensity $\Delta=100$; the overall clumping factor $\cb$ (which neglects 
the ionisation and temperature fields) scales with this value.
\begin{itemize}
\item In the absence of feedback, baryons follow dark matter, 
and $C_{100}=10.7$ as previously found in N-body simulations~\citep{ili07,rai11}.
\item Galactic outflows return star-forming gas into the IGM, boosting
$C_{100}$.  The increase is less than 30\% for any EUVB amplitude 
(including zero).
\item An ionising background pressurizes the diffuse IGM, suppressing
$C_{100}$ to 2.0.
\end{itemize}
In reality, the IGM's volume-averaged recombination rate depends on 
its density, ionisation, and temperature fields.  Taking this information
into account one piece at a time within the context of our fiducial
simulation, we find (still at $z=6$):
\begin{itemize}
\item The clumping factor of all baryons below our adopted threshold
density for star formation is $\cb=16.4$.
\item Averaging only over the ionised gas yields a much lower value
of $\cnenp=4.9$ because overdense gas is self-shielded.
\item Accounting for the temperature field further suppresses the 
recombination rate to $\cnenpobs=3.3$ because gas in the recent aftermath
of reionisation is somewhat hotter than the canonical $10^4$ K.  
\item The temperature-corrected clumping factor of ionised regions 
increases from $0.79$ at $z=15$ to 3.3 at $z=6$.  Equation~\ref{eqn:cfit} 
reproduces this evolution reasonably well throughout $z=15\rightarrow5$.
\item Our fiducial simulation may overestimate the minimum density of
neutral gas.  Correcting for
this may suppress $\cnenpobs$ by an additional factor of 12\%.  Hence 
our most realistic estimate of the ionisation- and temperature-corrected 
clumping factor at $z=6$ is that it lies within the range 2.7--3.3.
\end{itemize}

We have constructed an analytical reionisation model and tested how well 
different definitions of the clumping factor reproduce our numerical 
simulation's reionisation history.  We find that the clumping factor 
averaged over all IGM baryons $\cb$ significantly overestimates the 
IGM recombination rate because it treats self-shielded gas as if it
were optically thin.  Accounting for the ionisation field through 
$\cnenp$ improves agreement with the numerical simulation, although 
the recombination rate is still overestimated because the 
post-reionisation IGM is in reality slightly hotter than the canonical
$10^4$ K.  By contrast, using $\cnenpobs$ to 
account additionally for the IGM temperature field yields a monotonic 
reionisation history whose $\tes$ overestimates the numerical result by 
only 10\%.  We speculate that this small remaining difference owes at 
least partly to light-travel time effects that are missing from 
the analytical calculation although other effects may also contribute.

We have used our analytical model to estimate the photon
cost of reionising filaments, or regions with $\Delta=$1--50.  With 
our simulated emissivity history,
if some process (other than reionisation) heats the IGM at $z=19.5$ 
then the photon cost of reionisation is $\gamma/H=1.8$ photons per 
hydrogen atom.  Delaying the redshift at which the IGM is heated 
from $z=19.5$ to $z=9.0$ increases $\gamma/H$ to 2.9 because the 
clumping factor increases.  The difference between these numbers 
indicates that the Universe pays a price of $\approx1$ photon
per hydrogen in order to reionise its filaments.

The ionisation state of dense gas has observational implications
that merit further study.  Our simulations show that, in the presence
of self-shielding, outflows can boost the amount of gas at densities 
corresponding to the circumgalactic medium ($10 < \Delta < 1000$). 
If a significant fraction of this gas remains neutral, then it could
play two important roles in modulating the growth of the EUVB.  Within
halos, it could constitute a significant absorbing column, modifying
the fraction of ionising photons that escape the host halo.  This effect 
could mimic a redshift-dependent $\fesc$, which seems to be required by 
observations~\citep{ino06,fin11,haa12,kuh12}.  Additionally, gas that 
travels past the virial radius while retaining a self-shielded component 
could boost the abundance of Lyman limit systems, as suggested 
by~\citet{van12}.

A complementary motivation for improving our understanding of 
self-shielded gas is that it can be observed directly in the form of
damped Lyman $\alpha$ systems and low-ionisation metal absorbers.
The IGM's ionisation field is a major theoretical uncertainty hampering
efforts to interpret observations of low-ionisation metal ions.  For
example,~\citet{opp09} studied low-ionisation metal absorbers
through the use of homogeneous ionising backgrounds.  While this
approach may be appropriate for ions such as SiIV and CIV, it is
less appropriate for low-ionisation species such as OI and SiII
because it artificially ionises, heats, and smooths gas that should 
in reality be self-shielded.  Our current simulations allow dense gas to 
self-shield, hence they will yield improved predictions for the abundance 
of low-ionisation metal absorbers.  This will reduce the theoretical 
parameter space and increase the power of current observations.

\section*{Acknowledgements}
We thank the referee for a thoughtful and detailed report that
improved the draft.  We thank M.\ McQuinn and M.\ Prescott for helpful 
conversations.  KF thanks D.\ Godinez and J.\ Sivapalan for exploring 
the ability of galactic outflows to self-shield within our simulations.
Our simulations were run on the University of Arizona's Xeon cluster.  
Support for this work was provided by the National Science Foundation
through grant AST-0907998, the NASA Astrophysics Theory 
Program through grant NNG06GH98G, and grant number 
HST-AR-10647 from the SPACE TELESCOPE SCIENCE INSTITUTE, which is 
operated by AURA, Inc. under NASA contract NAS5-26555.  Support for 
this work, part of the Spitzer Space Telescope Theoretical Research 
Program, was also provided by NASA through a contract issued by the 
Jet Propulsion Laboratory, California Institute of Technology under 
a contract with NASA.  KF gratefully acknowledges support from NASA 
through Hubble Fellowship grant HF-51254.01 awarded by the Space 
Telescope Science Institute, which is operated by the Association 
of Universities for Research in Astronomy, Inc., for NASA, under 
contract NAS 5-26555.  SPO acknowledges support from NSF grant 
AST-0908480.

\onecolumn
\begin{deluxetable}{l|cccc}
\tabletypesize{\footnotesize}
\tablecolumns{5}
\tablewidth{0pt}
\tablecaption{Our suite of simulations.  The fiducial simulation is
indicated in bold. \label{table:sims}}
\tablehead{
\colhead{name} &
\colhead{$L$\tablenotemark{a}} &
\colhead{winds} &
\colhead{RT grid} &
\colhead{$M_{h\mbox{,min}}/\msun$\tablenotemark{b}}
}
\startdata
r6n256nWnRT   & 6 	& no   	& --  		& $1.4\times10^8$ \\
{\bf r6n256nWwRT16} & {\bf 6}	& {\bf no}  	& $\mathbf{16^3}$	& $ \mathbf{1.4\times10^8}$ \\
r6n256wWnRT   & 6	& yes	& --  		& $1.4\times10^8$ \\
r6n256wWwRT16 & 6	& yes	& $16^3$	& $1.4\times10^8$ \\
r6n256wWwRT32 & 6	& yes	& $32^3$	& $1.4\times10^8$ \\
r6n128wWwRT16 & 6	& yes	& $16^3$	& $1.1\times10^9$ \\
r3n128wWwRT8  & 3	& yes	& $8^3$		& $1.4\times10^8$ \\
r6n256nWHM01  & 6   	& no    & --            & $1.4\times10^8$ \\
\enddata
\tablenotetext{a}{in comoving $\hmpc$}
\tablenotetext{b}{virial mass of a halo with 100 dark matter and SPH particles.}
\end{deluxetable}

\appendix
\section{Approximations to the Eddington Tensor}\label{app:fedd}
Our method of solving the radiative transport equation involves using
a time-independent ray-tracing calculation to update the Eddington
tensors in a way that tracks the evolving emissivity and opacity
fields without becoming too computationally burdensome~\citep{fin09a}.
The radiation transfer simulations in Table~\ref{table:sims} use 
two approximations beyond those described in~\citet{fin09a}.

First, we smooth each element of the Eddington tensor field with 
a 27-cell cubical tophat filter.  This is necessary because the Eddington 
tensor can change rapidly at the positions of ionisation fronts, 
which can in turn cause numerical errors when the radiation field 
is updated.  We introduced this approach in~\citet{fin09a} but have
not previously used it within the context of cosmological simulations.
We stress that it does not materially degrade our spatial resolution 
because the smoothing occurs over the same length scale as the 
finite-difference stencil that we use to discretize the moments of 
the radiation transport equation.

Second, we update each cell's Eddington tensor less frequently as
it becomes optically thin.  Here, the assumption is that a region's 
Eddington tensor evolves rapidly as an ionisation front sweeps 
over it and more slowly afterwards.  We implement this idea as
follows: The code only updates a cell's Eddington tensor if the 
cell's photon number density changes by more than a factor $\fJ$.  
Whereas previously we set $\fJ=0.05$, we now allow $\fJ$ to
depend on the cell's optical depth $\tau = \chi \dxrt$, where
$\chi$ and $\dxrt$ are the local opacity and the cell size,
respectively.  In particular, $\fJ=0.05$ if $\tau\geq1$ and it
grows linearly from 0.05 to 1 as $\tau$ decreases from 1 to 0:
\begin{eqnarray}
\fJ = \left\{ 
\begin{array}{lc}
  0.05 & \tau \geq 1 \\
  1.0 - 0.95\tau & \tau < 1
\end{array} \right.
\end{eqnarray}
With this assumption, the Eddington tensor is updated frequently
before the cell is reionised and less frequently once the radiation
field has been established.

\section{Resolution Limitations}\label{app:res}

\begin{figure}
\centerline{
\setlength{\epsfxsize}{0.5\textwidth}
\centerline{\epsfbox{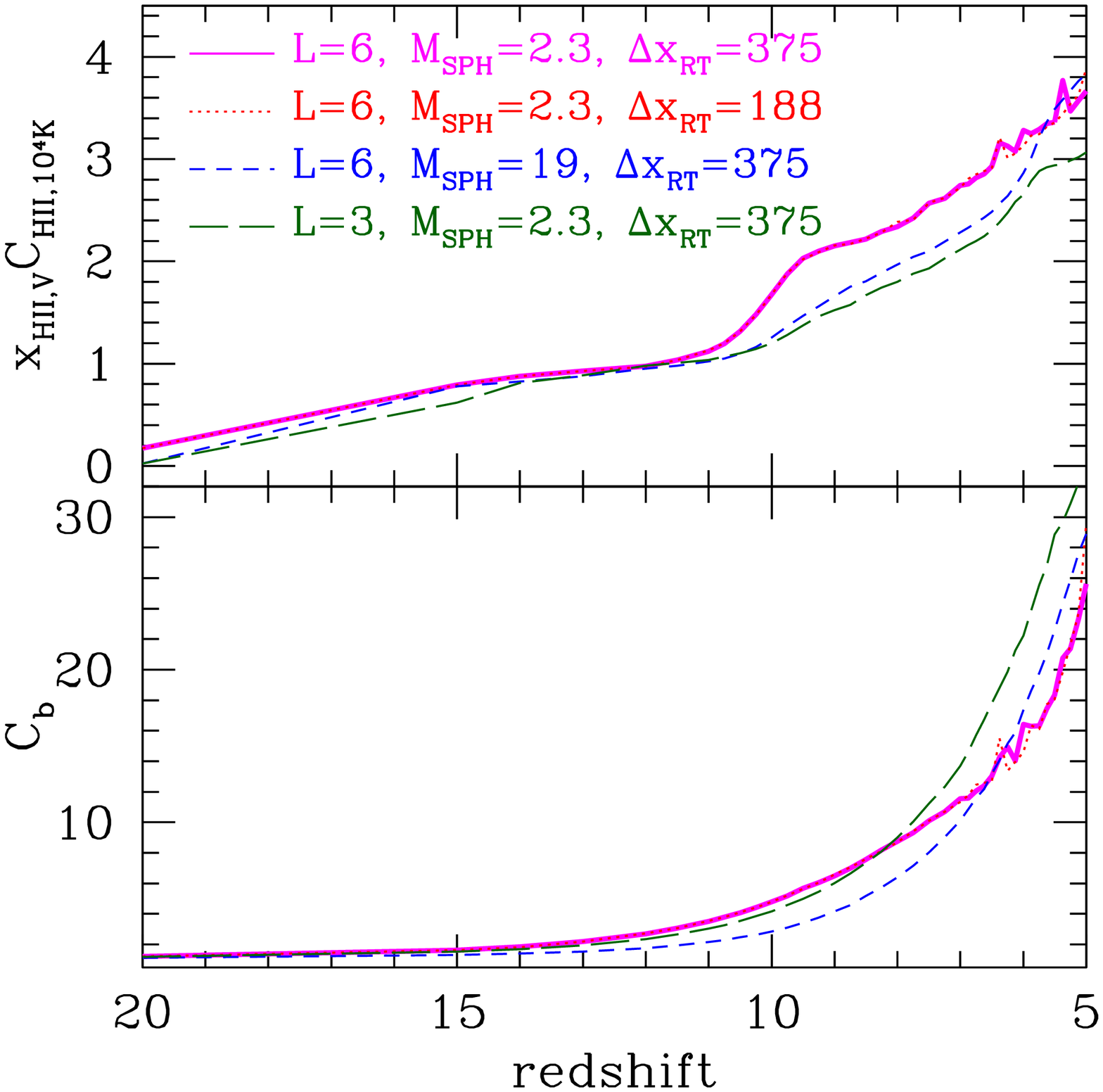}}
}
\caption{The impact of various resolution limitations on the 
temperature-corrected clumping factor of ionised gas.  The
simulation volumes' side lengths $L$ are in $\hmpc$, the
SPH particle masses are in $10^5\msun$, and the radiation
transport grid sizes are in comoving $\hkpc$.  Resolution
limitations are generally weak, indicating that our fiducial 
simulation volume (solid magenta) is reasonably converged 
numerically.
}
\label{fig:clumpres}
\end{figure}

In this section, we assess the sensitivity of our predicted clumping 
factors to three different flavors of resolution limitation: The 
simulation volume, baryonic mass resolution, and the spatial 
resolution of our radiation transport solver.  We evaluate 
convergence in Figure~\ref{fig:clumpres} by comparing the baryonic and
observational temperature-corrected clumping factors from 
different simulations with different numerical parameters but 
the same physical assumptions.

\subsection{Simulation Volume}
Our limited simulation volume could impact our results by modifying the 
relative contribution of voids and overdensities to the overall clumping 
factor~\citep{bol09,rai11}.  By delaying the growth of 
structure~\citep{bar04}, small volumes also change the dependence 
of ionisation and temperature on overdensity at fixed redshift.  Note
that we have mitigated some of these problems by tuning $\fesc(z)$ so 
that our simulations match a variety of observations (although not all 
observations; Section~\ref{sec:sims}), 
with the result that the simulations are \emph{representative} even if 
they are not \emph{converged}~\citep[see also][]{gne06}.  Nonetheless,
it is of interest to test our sensitivity to our limited volume.  The 
solid magenta and long-dashed green curves in Figure~\ref{fig:clumpres}
correspond to simulations that 
subtend different cosmological volumes at the same physical resolution, 
and their predicted temperature-corrected clumping factors of ionised 
gas (top panel) are reasonably close.  In detail, the simulation with 
the larger volume (solid magenta) predicts a slightly higher 
recombination rate at all times, with the gap growing to 20--30\% by 
$z=5$.  This difference owes entirely to the fact that, in a larger
volume, reionisation begins sooner, hence at any given redshift it
has proceeded farther from the voids into the overdense 
regions.  This is especially true in our simulations because our
self-shielding treatment effectively enforces an outside-in 
reionisation topology on small scales.  In other words, for the same 
dependence of ionisation fraction and temperature on overdensity, 
the predicted recombination rates would be converged.

Turning to the bottom panel, we see that the small-volume simulation
has a higher volume-averaged recombination rate $\cb$.  This may also 
be attributed to the delayed progress of reionisation: At any given
redshift, photoheating has smoothed the density 
fluctuations up to a lower density threshold in the small-volume 
simulation because the ionising background is weaker.  Consequently,
the clumping factor averaged over all IGM baryons (including the 
self-shielded ones) is higher.

This test does not account for fluctuations in the density field that
are on larger scales than our largest simulation volume, which is 
comparable to the typical size scale of ionised regions when the 
Universe is 50\% ionised~\citep{fur05}.  Similarly, it does not 
account for fluctuations in the EUVB on large ($>10\hmpc$) scales, 
which can in turn source large-scale fluctuations in the ionisation 
field~\citep{cro11}.  Future work at higher dynamic range will be
required to assess these limitations.

\subsection{Baryonic Mass Resolution}
The mass resolution of our SPH discretization dictates the amount of 
dense structure that can form, which in turn impacts the recombination 
rate by regulating the progress of reionisation.  The solid 
magenta and short-dashed blue curves in 
the top panel of Figure~\ref{fig:clumpres} illustrate these effects.  The
result of reducing the mass resolution at constant volume is qualitatively 
similar to reducing the simulation volume at constant resolution because
both delay reionisation.  In particular, the 
curve corresponding to the higher-resolution simulation (solid magenta) 
climbs systematically above its lower-resolution counterpart below $z=13$
because reionisation has proceeded into regions of higher overdensity
at a given redshift, boosting the mean recombination rate.  Following 
reionisation ($z<7$), the recombination rate in the lower-resolution 
simulation rejoins the prediction from the high-resolution simulation
because the predicted gas density distributions and ionising backgrounds
are similar.  Turning to the clumping factor of all baryons $\cb$ in the 
bottom panel, we see that the simulation with lower mass resolution has 
lower clumping prior to reionisation because the amount of matter that 
collapses into filaments and halos is lower at lower mass resolution.
Following reionisation, $\cb$ is higher at lower mass resolution
because its weaker ionising background does not penetrate as far into 
overdense regions.

It is important to note that varying the baryonic mass resolution 
changes the predicted clumping even if the spatial resolution of the 
radiation solver is unchanged.  This 
illustrates the benefit of solving the cooling and ionisation equations 
on the SPH particles rather than on the RT grid.  On the other hand, the
change is not large despite the order of magnitude increase in resolution,
indicating that the requirements for numerically resolving the clumping 
factor are not strict and that our simulations are therefore reasonably 
converged.

\subsection{Radiation Field Resolution}
A central benefit of modeling a spatially-resolved EUVB is the ability
to resolve self-shielding within overdense regions.  Once reionisation
is well under way, the boundaries of shelf-shielded regions dominate the 
IGM opacity and the clumping factor.  Increasing the radiation field's
resolution treats self-shielding more accurately, preventing overdense 
regions from becoming photoionised and smoothed.  While our
subgrid self-shielding prescription prevents the most overdense regions
from being photoionised, which limits the clumping factor, our radiation
solver's grid size is sufficiently large that it could still ionise
regions near the self-shielding threshold too efficiently; the result
would overestimate gas clumping.  Additionally, modeling the radiation 
field with a low spatial resolution artificially broadens ionisation
fronts, extending the time during which partially-ionised gas cools
through collisional excitation of neutral hydrogen and underestimating
the post-reionisation temperature.  Our fiducial simulation's spatial
resolution corresponds to an optical depth of 2 at the Lyman limit
and the mean density, hence this could affect our results~\citep{mir94}
In order to explore these possibilities, we compare the solid magenta 
and dotted red curves in 
Figure~\ref{fig:clumpres}, which show how the clumping factor evolves 
when the radiation field is discretized using our fiducial and twice
the fiducial resolutions, respectively.  The predictions are
nearly indistinguishable, indicating that the clumping factors are
insensitive to the resolution of our radiation transport solver.  This
resolution convergence is a clear demonstration of the power of our 
subgrid self-shielding prescription.

There are other numerical effects that we cannot consider directly using
convergence tests.  First, our use of a monochromatic radiation
solver likely results in underestimating the IGM temperature, particularly
in voids, because it does not capture spectral hardening~\citep{abe99}
or heating owing to absorption of high-energy photons by 
helium atoms~\citep{cia12}.
This would slightly suppress the IGM recombination rate.  On the other
hand, the inability of our hydrodynamic solver to resolve minihalos 
means that our simulations may underestimate the IGM recombination 
rate~\citep{hai01,sha04,ili07,cia06}.  Resolving the question of whether
numerical limitations cause us to over- or underestimate the clumping
factor will require simulations with significantly higher dynamic 
range.


\begin{thebibliography}{99}
\frenchspacing
\bibitem[Abel \& Haehnelt(1999)]{abe99} Abel, T., \& Haehnelt, M.~G.\ 1999, \apjl, 520, L13 
\bibitem[Barkana \& Loeb(1999)]{bar99} Barkana, R., \& Loeb, A.\ 1999, \apj, 523, 54 
\bibitem[Barkana \& Loeb(2002)]{bar02} Barkana, R., \& Loeb, A.\ 2002, \apj, 578, 1 
\bibitem[Barkana \& Loeb(2004)]{bar04} Barkana, R., \& Loeb, A.\ 2004, \apj, 609, 474 
\bibitem[Bolton \& Haehnelt(2007)]{bol07} Bolton, J.~S., \& Haehnelt, M.~G.\ 2007, \mnras, 382, 325 
\bibitem[Bolton \& Becker(2009)]{bol09} Bolton, J.~S., \& Becker, G.~D.\ 2009, \mnras, 398, L26 
\bibitem[Bolton et al.(2012)]{bol12} Bolton, J.~S., Becker, 
G.~D., Raskutti, S., et al.\ 2012, \mnras, 419, 2880 
\bibitem[Bouwens et al.(2011)]{bou11} Bouwens, R.~J., 
Illingworth, G.~D., Oesch, P.~A., et al.\ 2011, \apj, 737, 90 
\bibitem[Ciardi et al.(2006)]{cia06} Ciardi, B., Scannapieco, 
E., Stoehr, F., et al.\ 2006, \mnras, 366, 689 
\bibitem[Ciardi et al.(2012)]{cia12} Ciardi, B., Bolton, 
J.~S., Maselli, A., \& Graziani, L.\ 2012, \mnras, 423, 558 
\bibitem[Crociani et al.(2011)]{cro11} Crociani, D., 
Mesinger, A., Moscardini, L., \& Furlanetto, S.\ 2011, \mnras, 411, 289 
\bibitem[Dav{\'e} et al.(2006)]{dav06} Dav{\'e}, R., 
Finlator, K., \& Oppenheimer, B.~D.\ 2006, \mnras, 370, 273 (DFO06)
\bibitem[Dunlop et al.(2012)]{dun12} Dunlop, J.~S., McLure, 
R.~J., Robertson, B.~E., et al.\ 2012, \mnras, 420, 901 
\bibitem[Efstathiou(1992)]{efs92} Efstathiou, G.\ 1992, 
\mnras, 256, 43P 
\bibitem[Eisenstein \& Hu(1999)]{eis99} Eisenstein, D.~J., \& Hu, W.\ 1999, \apj, 511, 5 
\bibitem[Faucher-Gigu{\`e}re et al.(2008)]{fau08} 
Faucher-Gigu{\`e}re, C.-A., Lidz, A., Hernquist, L., 
\& Zaldarriaga, M.\ 2008, \apj, 688, 85 
\bibitem[Fernandez \& Shull(2011)]{fer11} Fernandez, E.~R., \& Shull, J.~M.\ 2011, \apj, 731, 20 
\bibitem[Fan et al.(2006)]{fan06} Fan, X., Strauss, M.~A., 
Becker, R.~H., et al.\ 2006, \aj, 132, 117 
\bibitem[Finkelstein et al.(2010)]{fin10} Finkelstein, S.~L., 
Papovich, C., Giavalisco, M., et al.\ 2010, \apj, 719, 1250 
\bibitem[Finlator et al.(2009a)]{fin09a} Finlator, K., 
{\"O}zel, F., \& Dav{\'e}, R.\ 2009, \mnras, 393, 1090 
\bibitem[Finlator et al.(2009b)]{fin09b} Finlator, K., 
{\"O}zel, F., Dav{\'e}, R., \& Oppenheimer, B.~D.\ 2009, \mnras, 400, 1049 
\bibitem[Finlator et al.(2011)]{fin11} Finlator, K., 
Dav{\'e}, R., {\&Ouml}zel, F.\ 2011, \apj, 743, 169 
\bibitem[Friedrich et al.(2012)]{fri12} Friedrich, M.~M., 
Mellema, G., Iliev, I.~T., \& Shapiro, P.~R.\ 2012, \mnras, 421, 2232 
\bibitem[Furlanetto et al.(2004)]{fur04} Furlanetto, S.~R., 
Zaldarriaga, M., \& Hernquist, L.\ 2004, \apj, 613, 1 
\bibitem[Furlanetto \& Oh(2005)]{fur05} Furlanetto, S.~R., \& Oh, S.~P.\ 2005, \mnras, 363, 1031 
\bibitem[Furlanetto \& Oh(2009)]{fur09} Furlanetto, S.~R., \& Oh, S.~P.\ 2009, \apj, 701, 94 
\bibitem[Gnedin \& Ostriker(1997)]{gne97} Gnedin, N.~Y., \& Ostriker, J.~P.\ 1997, \apj, 486, 581 
\bibitem[Gnedin(2000)]{gne00} Gnedin, N.~Y.\ 2000, \apj, 542, 
535 
\bibitem[Gnedin \& Fan(2006)]{gne06} Gnedin, N.~Y., \& Fan, X.\ 2006, \apj, 648, 1 
\bibitem[Gnedin et al.(2008)]{gne08} Gnedin, N.~Y., Kravtsov, 
A.~V., \& Chen, H.-W.\ 2008, \apj, 672, 765 
\bibitem[Gonz{\'a}lez et al.(2011)]{gon11} Gonz{\'a}lez, V.,
Labb{\'e}, I., Bouwens, R.~J., Illingworth, G., Franx, M.,
\& Kriek, M.\ 2011, \apjl, 735, L34
\bibitem[Grazian et al.(2011)]{gra11} Grazian, A., et al.\ 2011, \aap, 532, A33
\bibitem[Haardt \& Madau(2001)]{haa01} Haardt, F. \& Madau, P. 2001,
in Clusters of Galaxies and the High Redshift Universe Observed in X-rays, 64 (HM01)
\bibitem[Haardt \& Madau(2012)]{haa12} Haardt, F., \& Madau, P.\ 2012, \apj, 746, 125 
\bibitem[Haiman et al.(2001)]{hai01} Haiman, Z., Abel, T., 
\& Madau, P.\ 2001, \apj, 551, 599 
\bibitem[Hopkins et al.(2012)]{hop12} Hopkins, P.~F., 
Quataert, E., \& Murray, N.\ 2012, \mnras, 421, 3522 
\bibitem[Hui 
\& Gnedin(1997)]{hui97} Hui, L., \& Gnedin, N.~Y.\ 1997, \mnras, 292, 27 
\bibitem[Iliev et al.(2005)]{ili05} Iliev, I.~T., Shapiro, 
P.~R., \& Raga, A.~C.\ 2005, \mnras, 361, 405 
\bibitem[Iliev et al.(2007)]{ili07} Iliev, I.~T., Mellema, 
G., Shapiro, P.~R., \& Pen, U.-L.\ 2007, \mnras, 376, 534 
\bibitem[Inoue et al.(2006)]{ino06} Inoue, A.~K., Iwata, I., 
\& Deharveng, J.-M.\ 2006, \mnras, 371, L1 
\bibitem[Katz et al.(1996)]{kat96} Katz, N., Weinberg, D.~H., 
\& Hernquist, L.\ 1996, \apjs, 105, 19 
\bibitem[Kennicutt(1998)]{ken98} Kennicutt, R.~C. 1998, ApJ, 498, 541
\bibitem[Kohler et al.(2007)]{koh07} Kohler, K., Gnedin, 
N.~Y., \& Hamilton, A.~J.~S.\ 2007, \apj, 657, 15 
\bibitem[Komatsu et al.(2011)]{kom11} Komatsu, E., Smith, 
K.~M., Dunkley, J., et al.\ 2011, \apjs, 192, 18 
\bibitem[Kuhlen 
\& Faucher-Gigu{\`e}re(2012)]{kuh12} Kuhlen, M., \& Faucher-Gigu{\`e}re, C.-A.\ 2012, \mnras, 423, 862 
\bibitem[Mac Low \& Ferrara(1999)]{mac99} Mac Low, M.-M., \& Ferrara, A.\ 1999, ApJ, 513, 142 
\bibitem[Madau et al.(1999)]{mad99} Madau, P., Haardt, F., 
\& Rees, M.~J.\ 1999, \apj, 514, 648 
\bibitem[Madau et al.(2004)]{mad04} Madau, P., Rees, M.~J., 
Volonteri, M., Haardt, F., \& Oh, S.~P.\ 2004, \apj, 604, 484 
\bibitem[Martin(2005)]{mar05} Martin, C.~L.\ 2005, \apj, 621, 
227 
\bibitem[McLure et al.(2011)]{mcl11} McLure, R.~J., Dunlop, 
J.~S., de Ravel, L., et al.\ 2011, \mnras, 418, 2074 
\bibitem[McQuinn et al.(2007)]{mcq07} McQuinn, M., Lidz, A.,
Zahn, O., Dutta, S., Hernquist, L.,
\& Zaldarriaga, M.\ 2007, \mnras, 377, 1043
\bibitem[McQuinn et al.(2009)]{mcq09} McQuinn, M., Lidz, A., 
Zaldarriaga, M., et al.\ 2009, \apj, 694, 842 
\bibitem[McQuinn et al.(2011)]{mcq11} McQuinn, M., Oh, S.~P., 
\& Faucher-Gigu{\`e}re, C.-A.\ 2011, \apj, 743, 82 
\bibitem[Miralda-Escud{\'e} 
\& Rees(1994)]{mir94} Miralda-Escud{\'e}, J., \& Rees, M.~J.\ 1994, \mnras, 266, 343 
\bibitem[Miralda-Escud{\'e} et al.(2000)]{mir00} 
Miralda-Escud{\'e}, J., Haehnelt, M., \& Rees, M.~J.\ 2000, \apj, 530, 1 
\bibitem[Murray et al.(2005)]{mur05} Murray, N., Quataert, 
E., \& Thompson, T.~A.\ 2005, \apj, 618, 569 
\bibitem[Mu{\~n}oz \& Loeb(2011)]{mun11} Mu{\~n}oz, J.~A., \& Loeb, A.\ 2011, \apj, 729, 99 
\bibitem[Nestor et al.(2011)]{nes11} Nestor, D.~B., Shapley, 
A.~E., Steidel, C.~C., \& Siana, B.\ 2011, \apj, 736, 18 
\bibitem[Oesch et al.(2012)]{oes12} Oesch, P.~A., Bouwens, 
R.~J., Illingworth, G.~D., et al.\ 2012, arXiv:1201.0755 
\bibitem[Oh(2001)]{oh01} Oh, S.~P.\ 2001, \apj, 553, 499 
\bibitem[Oh \& Haiman(2003)]{oh03} Oh, S.~P., \& Haiman, Z.\ 2003, \mnras, 346, 456 
\bibitem[Oppenheimer 
\& Dav{\'e}(2008)]{opp08} Oppenheimer, B.~D., \& Dav{\'e}, R.\ 2008, \mnras, 387, 577 
\bibitem[Oppenheimer et al.(2009)]{opp09} Oppenheimer, B.~D., 
Dav{\'e}, R., \& Finlator, K.\ 2009, \mnras, 396, 729 
\bibitem[Pawlik \& Schaye(2009)]{paw09b} Pawlik, A.~H., \& 
Schaye, J.\ 2009, \mnras, 396, L46 
\bibitem[Pawlik et al.(2009)]{paw09a} Pawlik, A.~H., Schaye, 
J., \& van Scherpenzeel, E.\ 2009, \mnras, 394, 1812 
\bibitem[Powell et al.(2011)]{pow11} Powell, L.~C., Slyz, A., 
\& Devriendt, J.\ 2011, \mnras, 414, 3671 
\bibitem[Pritchard et al.(2010)]{pri10} Pritchard, J.~R., 
Loeb, A., \& Wyithe, J.~S.~B.\ 2010, \mnras, 408, 57 
\bibitem[Rai{\v c}evi{\'c} 
\& Theuns(2011)]{rai11} Rai{\v c}evi{\'c}, M., \& Theuns, T.\ 2011, \mnras, 412, L16 
\bibitem[Ricotti 
\& Ostriker(2004)]{ric04} Ricotti, M., \& Ostriker, J.~P.\ 2004, \mnras, 352, 547 
\bibitem[Schaerer(2003)]{sch03} Schaerer, D.\ 2003, \aap, 397, 527 
\bibitem[Schaye(2001)]{sch01} Schaye, J.\ 2001, \apj, 559, 507 
\bibitem[Schaye(2004)]{sch04} Schaye, J.\ 2004, \apj, 609, 
667 
\bibitem[Schaye et al.(2010)]{sch10} Schaye, J., Dalla 
Vecchia, C., Booth, C.~M., et al.\ 2010, \mnras, 402, 1536 
\bibitem[Shapiro et al.(2004)]{sha04} Shapiro, P.~R., Iliev, 
I.~T., \& Raga, A.~C.\ 2004, \mnras, 348, 753 
\bibitem[Shapley et al.(2006)]{sha06} Shapley, A.~E., 
Steidel, C.~C., Pettini, M., Adelberger, K.~L., 
\& Erb, D.~K.\ 2006, \apj, 651, 688 
\bibitem[Shull et al.(2012)]{shu12} Shull, J.~M., Harness, 
A., Trenti, M., \& Smith, B.~D.\ 2012, \apj, 747, 100 
\bibitem[Siana et al.(2010)]{sia10} Siana, B., Teplitz, 
H.~I., Ferguson, H.~C., et al.\ 2010, \apj, 723, 241 
\bibitem[Springel \& Hernquist(2003)]{spr03} Springel, V., \& Hernquist, L.\ 2003, \mnras, 339, 312
\bibitem[Springel(2005)]{spr05} Springel, V.\ 2005, \mnras, 364, 1105
\bibitem[Sutherland \& Dopita(1993)]{sut93} Sutherland, R.~S. \& Dopita, M. A.
1993, ApJS, 88, 253
\bibitem[Theuns et al.(2001)]{the01} Theuns, T., Mo, H.~J., 
\& Schaye, J.\ 2001, \mnras, 321, 450 
\bibitem[Trac \& Cen(2007)]{tra07} Trac, H., \& Cen, R.\ 2007, \apj, 671, 1 
\bibitem[Trac \& Gnedin(2009)]{tra09} Trac, H., \& Gnedin, N.~Y.\ 2009, arXiv:0906.4348
\bibitem[Trenti et al.(2010)]{tre10} Trenti, M., Stiavelli, 
M., Bouwens, R.~J., et al.\ 2010, \apjl, 714, L202 
\bibitem[Trenti et al.(2012)]{tre12} Trenti, M., Perna, R., 
Levesque, E.~M., Shull, J.~M., \& Stocke, J.~T.\ 2012, \apjl, 749, L38 
\bibitem[van de Voort et al.(2012)]{van12} van de Voort, F., 
Schaye, J., Altay, G., \& Theuns, T.\ 2012, \mnras, 421, 2809 
\bibitem[Venkatesan et al.(2001)]{ven01} Venkatesan, A., 
Giroux, M.~L., \& Shull, J.~M.\ 2001, \apj, 563, 1 
\bibitem[Weiner et al.(2009)]{wei09} Weiner, B.~J., Coil, 
A.~L., Prochaska, J.~X., et al.\ 2009, \apj, 692, 187 
\bibitem[Wong et al.(2008)]{won08} Wong, W.~Y., Moss, A., 
\& Scott, D.\ 2008, \mnras, 386, 1023 
\bibitem[Wyithe et al.(2008)]{wyi08} Wyithe, J.~S.~B., 
Bolton, J.~S., \& Haehnelt, M.~G.\ 2008, \mnras, 383, 691 
\end{thebibliography}
\end{document}